\documentclass[nofootinbib,preprintnumbers]{revtex4}
\usepackage{graphicx}
\usepackage{dcolumn}
\usepackage{amsmath,amssymb,epsfig}
\usepackage{paralist}
\usepackage{comment}
\usepackage{graphicx}
\usepackage{multirow}
\usepackage{wrapfig}

\usepackage[breaklinks=True,colorlinks=True, linkcolor=magenta,citecolor=blue,debug=True]{hyperref}

\renewcommand{\vec}[1]{\boldsymbol{\mathrm{#1}}}

\newcommand{\editnote}[2]{}

\allowdisplaybreaks

\begin{document}

\title{Searching for new physics in the solar system \\
with tetrahedral spacecraft formations}

\author{Slava G. Turyshev, Sheng-wey Chiow, Nan Yu}

\affiliation{
Jet Propulsion Laboratory, California Institute of Technology,\\
4800 Oak Grove Drive, Pasadena, CA 91109-0899, USA}

\date{\today}

\begin{abstract}

Tetrahedral configurations of spacecraft on unperturbed heliocentric orbits allow for highly precise observations of small spatial changes in the gravitational field, especially those affecting the gravity gradient tensor (GGT). The resulting high sensitivity may be used to search for new physics that could manifest itself via deviations from general relativistic behavior yielding a non-vanishing trace of the GGT. We  study the feasibility of recovering the trace[GGT]  with the sensitivity of  ${\cal O}(10^{-24}~{\rm s}^{-2})$ -- the level where some of the recently proposed cosmological models may have observable effects in the solar system. Specifically, we consider how a set of local measurements provided by precision laser ranging (to measure the inter-satellite ranges) and atom-wave interferometry (to correct for any local non-gravitational disturbances) can be used for that purpose. We report on a preliminary study of  such an experiment and on the precision that may be reached in measuring the trace[GGT], with the assumption of drag-compensated spacecraft by atom interferometer measurements.  For that, we study the dynamical behavior of a tetrahedral formation established by four spacecraft placed on nearby elliptical orbits around the Sun. We develop analytical models for the relevant observables and study the conditions for setting up an optimal tetrahedral configuration. We formulate the observational equations to measure the trace[GGT] relying only on the observables that are available within the formation, such as those based on the laser ranging and the Sagnac interferometry.  We demonstrate  that Sagnac observable is a mission enabling capability that allows to measure the angular frequency of the tetrahedral rotation with respect to an inertial reference frame with an accuracy that is much higher than that available from any other modern navigational techniques. We show that the quality of the science measurements is affected by the tetrahedron evolution, as its orientation and the shape change while the spacecraft follow their orbits. We present the preliminary mission and instrument requirements needed to measure the trace[GGT] to the required accuracy and thus demonstrate the feasibility of satisfying the stated science objective. 

\end{abstract}


\maketitle


\section{Introduction}

Precise measurements of the gravitational field, conducted through high-precision experiments within the solar system, present crucial opportunities for testing modified theories of gravitation. These tests could play a pivotal role in either validating or challenging these theories, providing essential insights into our understanding of gravitational forces. In classical physics, the Newtonian gravitational field is described by the gravitational Poisson equation. A key implication of this equation is that the gravitational gradient tensor (GGT) has a zero trace value in a vacuum environment. However, this is not the case in many modified theories of gravitation, which predict qualitatively different outcomes. For instance, in Yukawa-type modified gravity theories \cite{Berezhiani:2009,Gonzalez:2023,Atazadeh:2007,Alexander:2018,Benisty:2022}, as well as in Galileon theory \cite{Padilla:2010,Trodden:2011,Deffayet:2013,Chow:2009}, the trace of the GGT is non-vanishing. This variance from the Newtonian model is a fundamental aspect of these theories, motivates  search for such novel mechanisms in high-precision experiments in the solar system \cite{Turyshev-etal-2007}.

Among the plausible mechanisms to explain dark energy, modified gravity theories offer an intriguing deviation from Einstein's general theory of relativity (GR). To do that, some of such theories  introduce a screening mechanism that depends on the environmental density. Screening mechanisms in physics are fundamentally categorized into two types: those dependent on local mass densities and those that are not. Scalar field theories such as chameleon and symmetron are examples of the former. These theories exhibit a unique `thin shell effect' where only the outermost layer of a substantial object interacts with dark energy fields. This selective interaction causes the dark energy force to be predominantly confined to this thin outer layer, reducing its overall observable impact. As a result, gravitational forces typically overshadow dark energy interactions in most observable phenomena. Despite this, the presence of the short-range dark energy force could still be detected through precision experiments in laboratory settings \cite{Hamilton-etal:2015,Elder-etal:2016,Chiow-etal:2022}.

In contrast, the Vainshtein screening mechanism follows a distinct approach. This mechanism involves the Vainshtein scalar field, which is mediated by a nonlinear `galileon' field. The nonlinearity of this field's equation of motion is characterized by a coupling constant, $r_{\tt c}$. The galileon force behaves similarly to gravity, $1/r^2$, at large distances from matter (beyond a Vainshtein radius of several hundred parsecs) but diminishes much more gradually ($1/\sqrt{r}$ for a cubic galileon) when closer to matter. The enormity of the Vainshtein radius makes it impractical to test this scalar field in terrestrial laboratory settings. Currently, the experimental limits on $r_c$ are constrained by tests of gravity's inverse square law, including lunar laser ranging and the analysis of gravitational wave propagation \cite{Sakstein:2018,Creminelli-Vernizzi:2017,Crisostomi-Koyama:2018}. In the realm of cosmology, the galileon field is considered a theoretically robust candidate for explaining the dark energy field.

Unlike classical gravity theories that fail solar system tests, these screened theories can effectively mask their non-GR behavior in high-density regions, such as our solar system, thus providing an optimal testing ground \cite{Bartlett:2021}. 
By observing gravitational dynamics and interactions in such settings, one can tease out the subtle signatures of these modified gravity theories, thus broadening our understanding of gravitational physics beyond GR \cite{Sakstein:2018,Sakstein:2020}.

In particular, the cubic galileon model \cite{Andrews:2013}, with its Vainshtein screening mechanism, offers an alternative explanation to deviations from Newtonian gravity, distinct from chameleon and symmetron models. Notably, it does not exhibit the thin shell effects. Instead, it modifies Newtonian gravity over long distances. Within the Vainshtein radius, which is approximately 100 pc for the Sun, its gravitational force doesn't follow the inverse square law (ISL) and acts inversely proportional to the square root of the distance \cite{Adelberger:2003}. Therefore,  detection of an anomalous behavior can essentially be a test of the ISL. With the typical values of the gravitational gradients in the solar system evaluated to be  $GM_\odot/r^3\simeq 3.96 \times 10^{-14}~{\rm s}^{-2} ({\rm AU}/r)^3$, one would expect the effect due to cubic galileon model to be about a factor of $10^{10}$ times smaller, thus, setting a measurement sensitivity requirement of ${\cal O}(10^{-24}~{\rm s}^{-2})$ for a space-based test.  

Dark energy, if related to the cosmological constant $\Lambda$CDM model, has a unique characteristic wherein its pressure $p =-\rho$. Given the Universe's critical density of $\rho_{\tt c}\approx 0.85\times 10^{-26}~{\rm kg/m}^3$ \cite{Planck:2014} and dark energy contributing to $\sim 70$\% of this ($\Omega_{\tt DE}\simeq 0.7$), if U signifies the gravitational potential due to dark energy and the Laplacian, $\nabla^2$, indicates spatial variations, then the resulting contribution is $4\pi G \Omega_{\tt DE} \rho_{\tt c}\simeq -5.56 \times 10^{-36}~{\rm s}^{-2}$. This very small value underscores the subtle effect of dark energy on gravitational potential, confirming its broad yet delicate impact, particularly in driving the universe's accelerated expansion. The values associated with dark energy's effects, are extremely small and lie beyond the current detection capabilities of modern instruments. Given the present state of technology and understanding, these minute effects might remain unobservable in any direct manner. 

Recent interest in dark matter and dark energy detection has shifted towards the use of experimental search (as opposed to observational), particularly those enabled by atom interferometers (AI), as they offer a complementary approach to conventional methods. Situated in space, these interferometers utilize ultra-cold, falling atoms to measure differential forces along two separate paths, serving both as highly sensitive accelerometers and as potential dark matter detectors. In particular, it has been proposed \cite{Yu-etal:2018-Phase-I,Yu-etal:2019} that a tetrahedral formation of four interplanetary satellites placed on highly-elliptic heliocentric orbits may be used to measure the trace of the GGT of the Newtonian gravitational field at sufficient accuracy for direct detection of dark energy scalar field in the form of the proposed galileon model. Here we investigate the feasibility of such a mission, focusing on the tetrahedral orbits and its measurement precisions, where spacecraft are assumed drag-free. The drag-free reduction for each spacecraft could be achieved with the use of atom interferometers as accelerometers. We consider the achievable sensitivity of such an experimental concept and its implications for precision tests of gravity.

In this paper, we discuss the Gravity Probe and Dark Energy Detection Mission (GDEM), as proposed in \cite{Yu-etal:2018-Phase-I,Yu-etal:2019}. GDEM is designed to search for deviations from the canonical $1/r^2$ gravitational potential within the solar system. These spacecraft use high-precision laser ranging systems that allow simultaneous measurements of the GGT in four distinct directions. This design ensures the trace of the GGT vanish irrespective of the tetrahedron’s orientation. To optimize the conditions to detect the anticipated galileon signal that behaves $\propto 1/\sqrt{r}$, the GDEM spacecraft will be placed on nearby elliptic heliocentric orbits that will allow to sample the galileon field at various distances from the Sun. An elliptical orbit with varying distance from the Sun will allow the observation of such variation. This distance-dependent variation will reduce the systematics yielding stronger evidence for a GR violation, if observed.  

GDEM relies on AI to enable drag-free operations for spacecraft within a tetrahedron formation. We assume that AI can effectively measure and compensate non-gravitational forces, such as solar radiation pressure, outgassing, gas leaks, and dynamic disturbances caused by gimbal operations, as these spacecraft navigate their heliocentric orbits. We assume  that AI can compensate for local non-gravitational disturbances at an extremely precise level, down to differential accelerations of  $\sim1 \times 10^{-15}~\mathrm{m/s}^2/\sqrt{\rm Hz}$, the level that is technically feasible for AI, especially when deployed in space \cite{Sorrentino:2011,Knabe-eta:2022,Zhu-etal:2022,Premawardhana-etal:2023}. Our objective here is to ascertain the viability of using these technologies for a direct dark energy detection, especially to reach a targeted gradient sensitivity of $10^{-24}~{\rm s}^{-2}$ over a 3-year period, which matches the expected galileon signal at a distance equivalent to 1 AU from the Sun. 

As we shall see, it is possible to investigate this particular prediction at very high accuracy using a configuration of satellites in heliocentric orbits. These satellites would be flying in close formation, typically separated by a few thousand kilometers or less, as they follow individual, undisturbed eccentric orbits around the Sun. Relying solely on observations that can be performed within the constellation (and specifically not relying on precision astrometry or high accuracy radio navigation, as neither methods are sufficiently accurate for this experiment) it is possible to reconstruct the trace of the GGT. For this, we can utilize intersatellite range data and accelerations in the form of corresponding second time derivatives. There are, however, some difficulties that must be overcome. We recognize that the tetrahedron may also undergo rotation, which implies that the computed accelerations must be decoupled from pseudo-accelerations implied by a rotating reference frame. In the absence of an accurate external reference, these accelerations must be measured in-constellation. To allow for such measurements, we introduce a generalized Sagnac-type observable, which, in principle, can be measured at the required accuracy using the same laser ranging equipment (perhaps with modest modifications) that is already available on board for range measurements.

This paper is organized as follows: In Sec.~\ref{sec:ggravten}, we develop an analytical formalism that allows to estimate the value of the trace of the GGT using the data from four spacecraft that move on nearby elliptic orbits, thus forming a tetrahedral formation. We introduce a coordinate frame suitable for representing observations made within the constellation, and discuss the generalized Sagnac observable that allows us to account for the rotation of this frame.
In Sec.~\ref{sec:eccentric-orb}, we offer an analytical discussion of motion along a set of idealized, nearly identical eccentric orbits around the Sun.  We also report on a numerical simulation that we employed to verify and validate our results. We summarize our results in Sec.~\ref{sec:conclude} and discuss next steps. For convenience, we put some technically relevant material in the Appendices. Appendix~\ref{sec:partials} discusses relevant partial derivatives needed to develop linear perturbation of elliptic Keplerian orbits. 

\section{Measurements of the GGT with a tetrahedral formation}
\label{sec:ggravten}

Spacecraft formations and/or distributed space systems allow measurements not possible with single spacecraft. For fundamental physics missions, observable field parameters related to the gravitational field may exhibit subtle variations in space and time. Thus, understanding of the processes within the field requires accurate and precise observations and analysis of both the temporal and spatial variations. 

A tetrahedron formation has the advantage since four spacecraft with variable separations, are the minimum needed to resolve a three-dimensional structure, at least to the lowest order in the physical field gradients. The tetrahedral configuration is a special geometric arrangement that has been proposed for multiple spacecraft formations, primarily due to its inherent stability and symmetry properties. When discussing spacecraft in such configurations, they can be seen as vertices of a tetrahedron moving in space. Here we will consider the relevant spacecraft dynamics.

\subsection{Spacecraft dynamics in nearby orbits}

We consider a constellation of four spacecraft that move around the Sun in a close tetrahedral formation. As seen from the solar solar system's barycentric coordinate reference system (BCRS), the motion of a $j$-th spacecraft under the gravitational attraction from the Sun,  is governed by the classical equations of motion, given as
{}
\begin{equation}
\ddot {\vec R}_j=\vec \nabla_j U+\vec p_j,
\label{eq:eq-m}
\end{equation}
where ${\vec R}_j$ is the position of the spacecraft in the BCRS, $U$ is the solar gravitational potential, and $\vec p_j$ are the accelerations associated with other forces (both of gravitational and non-gravitational nature), spacecraft on-board disturbances, or control inputs. 
On the other hand, since we have the basic assumption on the drag-free spacecraft behavior, then the $\vec p_j$ would denote the unknown forces that we are interested to detect.

Our objective here is to investigate the architecture and the anticipated  sensitivities in measuring the GGT provided by a tetrahedral spacecraft configuration. For that, it is sufficient to consider only the non-relativistic Newtonian approximation to GR, which, in this case, collapses to 
{} 
\begin{equation}
\nabla^2 U=4\pi G\rho,
\label{eq:ENeq}
\end{equation}    
where $\rho$ is the density of matter within the solar system with $U$ being the resulting gravitational potential.

We consider the solar gravitational potential $U(\vec r)$, which, accounting for the axial symmetry around the rotation axis, may be given as below
{}
\begin{eqnarray}
U(\vec r)&=&\frac{\mu_\odot}{r}\Big[1-\sum_{n=1}^\infty\Big(\frac{R_\odot}{r}\Big)^{2n}J_{2n} P_{2n}(\cos\theta)\Big],
\label{eq:pot_w_0sh}
\end{eqnarray}
where $\mu_\odot=GM_\odot$, $J_{2n}$ are the solar gravitational moments that describe the rotation-induced deviation of the Sun’s outer gravitational potential
 from a spherical configuration, $P_{2n}$ are the Legendre polynomials, and $\theta$ is the colatitude or angle with respect to the rotation axis. With  (\ref{eq:pot_w_0sh}), we evaluate the GGT at $\sim 1$~AU from the Sun:
{}
\begin{equation}
\vec T \equiv  \nabla_a \nabla_bU=
  \begin{bmatrix}
\partial^2_{11}U & \partial^2_{12}U & \partial^2_{13}U\\[4pt]
\partial^2_{21}U& \partial^2_{22}U & \partial^2_{23}U\\[4pt]
\partial^2_{31}U& \partial^2_{31}U& \partial^2_{33}U\\
 \end{bmatrix}= -
\frac{\mu_\odot}{r^3}\Big(\delta^{ab}-3n^a n^b\Big)+{\cal O}\Big(2.84\times 10^{-25}~{\rm s}^{-2}\Big),
\label{eq:eq-ocsisws}
\end{equation}
where $n^a\equiv \vec n=\vec r/r$ and the error term is due to the solar gravitational quadrupole, $J_2$. Thus, for the target sensitivity of GDEM of $\sim 10^{-24}~{\rm s}^{-2}$ in measuring the GGT, it is sufficient  to treat the solar gravity field to be spherically-symmetric.\footnote{The order term ${\cal O}(...)$ indicates the largest omitted terms in the expression.}

Laplace equation (\ref{eq:ENeq}) says that $\vec T$ is trace-free (in vacuum and in the absence of other forces and fields.) This turns out to be the most powerful  check for the presence of new physical laws in the solar system. The measurement systems prescribed for GDEM involve assessing GGT between spacecraft pairs connected by laser ranging interferometers across considerable distances, important for reaching the required sensitivity. While similar to the inter-spacecraft laser ranging in LISA, GDEM's system is tailored for accuracy over these specific distances.

To describe the spacecraft dynamics within the constellation, we need to choose a reference point which could be either the formation's center or one of the satellites. With this choice, the $j$-th spacecraft position may be given as
{}
\begin{equation}
\vec R_j =\vec R_{\tt c}+\vec r_j, \qquad {\rm with} \qquad \ddot {\vec R}_{\tt c}=\vec \nabla_{\tt c} U,
\label{eq:pos-tr}
\end{equation}
where $\vec R_{\tt c}$ and $\vec r_j$ are the BCRS position of the  reference center and the relative position of the $j$-th spacecraft with respect to that reference center, respectively.

As follows from (\ref{eq:eq-m})--(\ref{eq:pos-tr}), the spacecraft motion with respect to the reference center is governed by  the equation
{}
\begin{equation}
\ddot {\vec r}_j=\Big(\vec \nabla_j U-\vec \nabla_{\tt c} U\Big)+\vec p_j.
\label{eq:rel-eq-m-c}
\end{equation}

Considering the solar gravitational potential $U$ in the form of (\ref{eq:pot_w_0sh}), we see that after the spherically-symmetric monopole term, the next largest contribution comes from the quadrupole term that is characterized by the solar quadrupole moment known to be $J_2=2.21\times 10^{-7}$ \citep{Mecheri-Meftah:2021}. A spacecraft in a heliocentric orbit with a semi-major axis of $a\simeq 1$~AU, would experience an acceleration due to the solar quadrupole of $a_{J_2}\sim 4.25\times 10^{-14}~{\rm m/s}^2$. For a pair of spacecraft separated by $r_{ij}\sim 10^3$~km, this would result in the differential acceleration of $\delta a_{ij}\simeq a_{J_2}(r_{ij}/{\rm 1\, AU})\sim 2.84\times 10^{-19}~{\rm m/s}^2$, corresponding to a gravity gradient signal of $\delta a_{ij}/r_{ij}\sim 2.84\times 10^{-25}~{\rm s}^{-2}$, which is negligible for our purposes. The next term is due to $J_4=-4.25\times 10^{-9}$ which would result in a signal that is $10^6$ times smaller than that of $J_2$ and thus may be neglected, so as the higher order terms. Clearly, the solar gravity oscillations \cite{Gough:1995}, which are several orders smaller than the primary quadrupole $J_2$ term, are going to make even smaller contributions to the trace of the GGT, thus are also negligible. Therefore, in (\ref{eq:pot_w_0sh}) it is sufficient to keep only the solar monopole, effectively, treating the sun as a point mass.
 
One may also consider relativistic contributions to the equations of motion of the spacecraft (\ref{eq:rel-eq-m-c}). Note that a typical Schwarzschild acceleration for the orbit with a semi-major axis of  $a\simeq1$~AU behaves as:
{}
\begin{eqnarray}
\ddot {\vec r}_{\rm GR}&=&\frac{\mu_\odot}{c^2r^3}\bigg\{\Big[2(\beta+\gamma)\frac{\mu_\odot}{r}-\gamma\dot{\vec r}^2\Big]\vec r+2(1+\gamma)({\vec r}\cdot\dot{\vec r})\dot{\vec r}\bigg\}\simeq 4.12 \times 10^{-10}~{\rm m/s}^2.
\label{eq:rel_eqmot_S}
\end{eqnarray}

Acceleration (\ref{eq:rel_eqmot_S}) translates in the differential acceleration between the vehicles of $\delta a^{\tt GR}_{ij}=\ddot {\vec r}_{\rm GR} (r_{ij}/1{\rm \, AU})=2.76 \times 10^{-15}~{\rm m/s}^2$ and the gravity gradient signal of $\delta a^{\tt GR}_{ij}/r_{ij}\simeq 2.76 \times 10^{-21}~{\rm s}^{-2}$, which is large and must be accounted. Thus, although here we are concerned only with the Newtonian gravity field and its contribution to the GGT, any future developments must consider the general relativistic terms in (\ref{eq:rel_eqmot_S}). That is in addition to the fact that the GGT must also be fully relativistic relying on the geodesic deviation equation of the relativistic space-time.

As a result, using only the monopole term in (\ref{eq:pot_w_0sh}), we may present the solar gravitational potential at the position of spacecraft $j$ as $U(\vec R_j)={\mu_\odot}/{R_j}$ and, while treating the constellation to be compact, so that $|\vec r_j|\ll |\vec R_{\rm c}|$, we expand the difference of the gravity gradients in (\ref{eq:rel-eq-m-c}) in terms of the small parameter $r_j/R_{\rm c}$, which yields:
{}
\begin{align}
(\vec \nabla_j U-\vec \nabla_{\rm c} U)^a&\simeq - \frac{\mu_\odot}{R_{\rm c}^3}\Big(\delta^{ab}-3 n^a_{\rm c} n^b_{\rm c}\Big) r^b_j - \frac{15 \mu_\odot}{2R_{\rm c}^4}{\cal Q}^{<abc>}r^b_j r^c_j - \frac{35 \mu_\odot}{2R_{\rm c}^5}{\cal Q}^{<abcd>}r^b_j r^c_jr^d_j +
{\cal O}\Big(r^4_j/R^6_{\rm c}\Big),
\label{eq:grad-diff}
\end{align}
where $n^a_{\tt c}\equiv\vec n_{\tt c}={\vec R_{\tt c}}/{R_{\tt c}}$ in the heliocentric position of the reference center chosen within the spacecraft formation.\footnote{The reference center may be associated  with one of the spacecraft within the formation or taken to be at the formations's mesocenter.} Also, the Einstein summation convention was used with indices $a,b=1,2,3$.
Also, ${\cal Q}^{<abc>}$ and ${\cal Q}^{<abcd>}$ are the Cartesian symmetric-trace free (STF) multipole coefficients \cite{Turyshev-Toth:2022} representing the gravitational tidal forces acting between  spacecraft $j$ and the formation center:
{}
\begin{eqnarray}
{\cal Q}^{<abc>}&=&\Big\{n_{\rm c}^a n_{\rm c}^b n_{\rm c}^c-{\textstyle\frac{1}{5}}\delta^{ab} n_{\rm c}^c-{\textstyle\frac{1}{5}}\delta^{ac}n_{\rm c}^b-{\textstyle\frac{1}{5}}\delta^{bc}n_{\rm c}^a\Big\},
\label{eq:sp-Tabc}\nonumber\\
{\cal Q}^{<abcd>}&=&\Big\{n_{\rm c}^an_{\rm c}^bn_{\rm c}^cn_{\rm c}^d-{\textstyle\frac{1}{7}}\Big(n_{\rm c}^an_{\rm c}^b\delta^{cd}+n_{\rm c}^a n_{\rm c}^c\delta^{bd}+n_{\rm c}^an_{\rm c}^d\delta^{bc}+n_{\rm c}^bn_{\rm c}^c\delta^{ad}+n_{\rm c}^bn_{\rm c}^d\delta^{ac}+n_{\rm c}^cn_{\rm c}^d\delta^{ab}\Big)+
\nonumber\\
&& \hskip 50pt +\,{\textstyle\frac{1}{35}}\Big(\delta^{ab}\delta^{cd}+\delta^{ac}\delta^{bd}+\delta^{ad}\delta^{bc}\Big)\Big\}.
\label{eq:sp-Tabcb}
\end{eqnarray}

We estimate the magnitudes of the terms present in (\ref{eq:grad-diff}): i) the term linear in $\propto r_i$ represents a differential acceleration of $\delta_1 a_{ij}\simeq 7.93\times 10^{-8}~{\rm m/s}^2$, yielding a GGT signal of $\delta_1 a_{ij}/r_{ij}\simeq 7.93\times 10^{-14}~{\rm s}^{-2}$, which is the signal of primary concern for GDEM (namely, we need to be able to measure and remove this GGT signal which is much larger than any signals that we are aiming to detect),  ii) the second term, $\propto r_i^2$, is responsible for an acceleration of $\delta_2 a_{ij}\simeq 7.95\times 10^{-13}~{\rm m/s}^2$, yielding a GGT contribution of $\delta_2 a_{ij}/r_{ij}\simeq 7.95\times 10^{-19}~{\rm s}^{-2}$, and finally, iii) the magnitude of the last term, $\propto r_i^3$, was evaluated to be  $\delta_3 a_{ij}\simeq 7.09\times 10^{-18}~{\rm m/s}^2$ and the corresponding GGT signal $\delta_3 a_{ij}/r_{ij}\simeq 7.09\times 10^{-24}~{\rm s}^{-2}$. Note that the first omitted terms ${\cal O}(r^4_j/R^6_{\rm c})$ or $\propto r_i^4$ would result in the acceleration contribution of $\delta_4 a_{ij}\simeq  1.18\times 10^{-23}~{\rm m/s}^2$  and the corresponding GGT signal of $\delta_4 a_{ij}/r_{ij}\simeq 1.18\times 10^{-29}~{\rm s}^{-2}$, that are negligible for our purposes. Therefore, to satisfy the science objectives of the GDEM mission, one needs to keep in the model all the terms present in (\ref{eq:grad-diff}).

To simplify the model development, we extend the definition of $\vec p_j$ from (\ref{eq:eq-m}) by adding to it relativistic terms (\ref{eq:rel_eqmot_S}) and those due to the tidal forces present in (\ref{eq:grad-diff}), thus defining the aggregate force term, $\vec f_j$,  as below
 {}
\begin{equation}
 f^a_j= p^a_j+\ddot { r}^a_{\rm GR}- \frac{15 \mu_\odot}{2R_{\rm c}^4}{\cal Q}^{<abc>}r^b_j r^c_j - \frac{35 \mu_\odot}{2R_{\rm c}^5}{\cal Q}^{<abcd>}r^b_j r^c_jr^d_j +{\cal O}\big(r^4_j/R^4_{\rm c}\big),
\label{eq:eq-ff}
\end{equation}
where indices from the first part of Latin alphabet $a,b,c,d$ denote vector components (with the Einstein summation convention used), while indices from the second part of Latin alphabet $i,j$ are used to denote a particular spacecraft. 

Given the fact that the terms $\ddot {\vec r}_{\rm GR}$ and those due to the tidal forces ${\cal Q}^{<abc>}$ and ${\cal Q}^{<abcd>}$ are small, one can directly account for their presence during data analysis as their magnitudes and behavior will be rather well-known.

Next, we observe that the first term in (\ref{eq:grad-diff}) is identical to the GGT, $\vec T$, given by  (\ref{eq:eq-ocsisws}). As a result, using  (\ref{eq:grad-diff}) and (\ref{eq:eq-ocsisws}), allows us to re-write (\ref{eq:rel-eq-m-c}) as below
{}
\begin{equation}
\ddot { r}^a_j=- \frac{\mu_\odot}{R_{\rm c}^3}\Big(\delta^{ab}-3 n^a_{\rm c} n^b_{\rm c}\Big) r^b_j+ f^a_j \qquad
\Rightarrow
\qquad
\ddot {\vec r}_j=\vec T_{\rm }\cdot \vec r_j+\vec f_j,
\label{eq:eq-bc2rs}
\end{equation}
which is our fundamental result providing us with the observational equation needed to determine the GGT. 

Equation (\ref{eq:eq-bc2rs}) describes the motion of a spacecraft with the constellation relative to the formation center from  the standpoint of the BCRS. However, our measurements are going to be conducted by the instruments placed on the orbiting spacecraft (i.e., atomic interferometers, laser ranging transceivers, and also the Sagnac interferometers). Furthermore, we will mostly rely on the relative measurements between spacecraft within the constellation. For a proper description of these measurements, we need to introduce an  orbital reference frame and appropriate coordinate system. Such a system is needed to describe the relative dynamics between the spacecraft and the relevant measurements as well as to transform the results between the orbital frame and the BCRS, if needed.

\subsection{Relevant orbital coordinate systems}
 \label{sec:loc-tetra}

\begin{figure}
\includegraphics[]{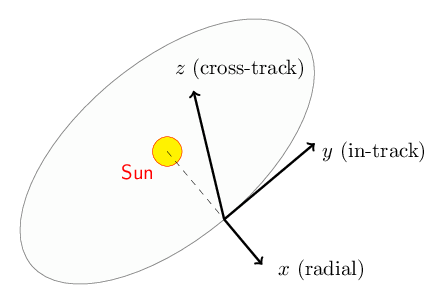}
\caption{\label{fig:RIC} The Radial--In-track--Cross-track (RIC) coordinate frame for a satellite or constellation in heliocentric orbit.}
\end{figure}

A natural coordinate frame for describing the relative motion of a spacecraft with respect to a reference center is the radial, in-track, cross-track (RIC) frame shown in Fig.~\ref{fig:RIC}.  This is a non-inertial frame that moves with the reference center. Assuming that position $\vec r$ and velocity $\dot{\vec r}$ of the reference center are known, evaluated at the center, the fundamental directions of this frame are (see more details in (\ref{eq:axis}))
{}
\begin{eqnarray}
\vec e_{\tt R}&=&\frac{\vec r}{r}, \quad~~ 
 \vec e_{\tt C}=\frac{\big[\vec r\times \dot{\vec r}\big]}{\big |\big[\vec r\times \dot{\vec r}\big]\big |}, 
 \quad~~ 
 \vec e_{\tt I}=\big[\vec e_{\tt C}\times \vec e_{\tt R}\big], \quad~~
  \label{eq:axis0}
\end{eqnarray}
with the unit vector $\vec e_{\tt R}$ pointing radially outward from Sun's center and $\vec e_{\tt I}$ is in the in-track direction along increasing true anomaly. This right-handed orthogonal reference frame is completed with $\vec e_{\tt C}$, pointing in the cross-track direction.

As will be shown in Sec.~\ref{sec:eccentric-orb}, the fundamental vectors describing position of the $i$-th spacecraft  $\vec r_i =(x_i,y_i,z_i)$ can be expressed in  the RIC frame (\ref{eq:axis0}) as
{}
\begin{eqnarray}
 \vec r_i &=& x_i\vec e_{\tt R}+y_i\vec e_{\tt I}+z_i\vec e_{\tt C},
\label{eq:eq-rho}\\
\vec r&=&  r\vec e_{\tt R},
\label{eq:eq-Rfc}
\qquad 
{\vec \omega}_{\tt RIC} = \dot{\theta}\vec e_{\tt C},
\label{eq:eq-om0}
\end{eqnarray}
where $\dot\theta$ is the angular velocity of the formation center (see relevant discussion in Sec.~\ref{sec:orb-pert}, especially after (\ref{eq:axis}).)

The RIC frame, also known as the local orbital frame, is a key coordinate system employed to represent the relative motion between satellites in the same orbit or between a satellite and its intended orbit. This system plays a crucial role in satellite activities such as formation flying, proximity operations, and relative navigation.  The versatility of the RIC frame is highlighted in satellite control and maneuvering, enabling adjustments in positioning to regulate ground track timings or to modify the satellite's orbital plane.

However, given the set of the laser ranging observables (which are the main observable of the GDEM mission), the RIC coordinate system is not the most convenient for the purposes of our experiment. As defined by (\ref{eq:eq-rho})--(\ref{eq:eq-om0}), this coordinate system (see Sec.~\ref{sec:eccentric-orb} for details)  relies on the knowledge of spacecraft positions in the BCRS. Such information will be provided by the NASA Deep Space Network (DSN)\footnote{See, \url{https://www.jpl.nasa.gov/missions/dsn}.} tracking in combination with the data from on-board star-trackers, which are not the most  precise set of measurements available.  Furthermore, our experiment relies on laser ranging that measures only the inter-satellite ranges $r_{ij}=|\vec r_{ij}|$. Thus, another coordinate system is needed, one that would allow expressing all the quantities of interest with respect to the length measurements  of the edges of the tetrahedron formed by four point masses in elliptical heliocentric orbits  \cite{Laus-Selig:2020}. Such a system may only be introduced on the tetrahedron itself.

\begin{figure}
\includegraphics[]{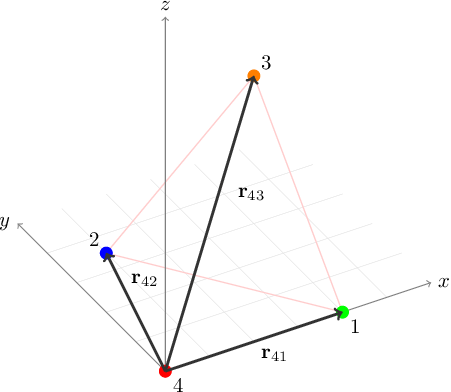}
\caption{\label{fig:TCS} The tetrahedral coordinate system (TCS) frame, constructed with satellite 4 as the origin. Note that the TCS frame is rotated by $\vec \omega_{\tt TCS}$ with respect to the RIC frame, as shown by (\ref{eq:angl-vel}).}
\end{figure}

Therefore, one needs to define the reference frame associated not only with one of the vertices of the tetrahedron, but with the tetrahedron itself, so that we could fully benefit from laser metrology that will provide highly precise measurements of all six edges of the tetrahedron. For that, we introduce an orthonormal coordinate system, associated with vertex \#4 of the tetrahedron (thus, the position of  spacecraft \#4 is $\vec r_4 =(0,0,0)$) (as shown in Fig.~\ref{fig:TCS}), as follows:
{}
\begin{align}
\vec e_x=\vec n_{41}, \qquad
\vec e_z=\frac{[\vec n_{41}\times \vec n_{42}]}{|[\vec n_{41}\times \vec n_{42}]|}, \qquad
\vec e_y=[\vec e_z\times\vec e_x],
 \label{eq:dg25da2=}
\end{align}
where $\vec r_{ij}=\vec r_{j}-\vec r_{i}$ and $\vec n_{ij}=\vec r_{ij}/r_{ij}$.

In this coordinate system, positions of each of the other three spacecraft forming the tetrahedron are given as:
{}
\begin{eqnarray}
\vec r_{41} &=& r_{41}\big(1,0,0\big)\equiv r_{41}\vec n_{41},
\label{eq:r_coord1}\\
\vec r_{42} &=& r_{42}\big(\cos\alpha_{12},\sin\alpha_{12},0\big)\equiv r_{42}\vec n_{42},\\
\vec r_{43} &=& r_{43}\big(\cos\alpha_{13},\sqrt{\sin^2\alpha_{13}-\sin^2\beta_3},\sin\beta_3\big)\equiv r_{43}\vec n_{43},
\label{eq:r_coord3}
\end{eqnarray}
where $\alpha_{12}$ and  $\alpha_{13}$ are the angles between the edges $\{41\}$ and $\{42\}$ and between those $\{41\}$ and $\{43\}$, correspondingly, and are given as
{}
\begin{eqnarray}
\cos \alpha_{12} &=& \frac{r^2_{41}+r^2_{42}-r^2_{12}}{2r_{41}r_{42}},
\qquad \sin \alpha_{12} =\sqrt{1-\cos^2 \alpha_{12} },
\label{eq:angles1}\\
\cos \alpha_{13} &=& \frac{r^2_{41}+r^2_{43}-r^2_{13}}{2r_{41}r_{43}},
\qquad \sin \alpha_{13} =\sqrt{1-\cos^2 \alpha_{13} }.
\label{eq:angles2}
\end{eqnarray}

Also, $\beta_{3}$ is the angle between $\vec n_{43}$ and the plane formed by $\vec n_{41}$ and $\vec n_{42}$. This angle is given as
{}
\begin{eqnarray}
\sin \beta_3 &=& (\vec n_{43}\cdot[\vec n_{41}\times \vec n_{42}])=\pm \frac{6V}{r_{41}r_{42}r_{43}},
\label{eq:angles3}
\end{eqnarray}
where $V$ is the volume of the tetrahedron, which is given as
{}
\begin{eqnarray}
V &=& {\textstyle\frac{1}{12}}\Big(4r^2_{41}r^2_{42}r^2_{43}-r_{41}^2 u^2-r_{42}^2v^2-r_{43}^2w^2+uvw\Big)^\frac{1}{2},\\
u&=& r_{42}^2+r_{43}^2-r_{23}^2, \qquad
v= r_{41}^2+r_{43}^2-r_{13}^2,\qquad
w=r_{41}^2+r_{42}^2-r_{12}^2.
\label{eq:vol-tetra}
\end{eqnarray}

The definition above yields the local coordinate system defined at the tetrahedron that we shall call the tetrahedron coordinate system (TCS). As one may see, the position of every spacecraft in this coordinate system  (\ref{eq:r_coord1})--(\ref{eq:r_coord3}), is now expressed in terms on the lengths of the six edges of the tetrahedron, as evidenced by (\ref{eq:angles1})--(\ref{eq:vol-tetra}).  This justifies the choice for the TCS, as it allows one to relate the elastic behavior of the tetrahedron to the set of available laser ranging measurements that will provide highly precise information on the  length of the six edges of the tetrahedron.

\subsection{Relative spacecraft dynamics in the TCS}
\label{sec:tcs-ric}

The two  coordinate systems discussed above, namely TCS, defined by (\ref{eq:dg25da2=}) and the Radial-In-track-Cross-track (RIC) system that is nominally used to describe proximity operations in an orbital frame introduced by (\ref{eq:eq-rho})--(\ref{eq:eq-om0}) and will be discussed in Sec.~\ref{sec:orb-pert}, have the same origin and are related by rotation
  {}
\begin{equation}
\vec e_{\tt TCS} = \vec R_{\tt TCS}  \vec e_{\tt RIC},
\label{eq:rot}
\end{equation}
where $ \vec R_{\tt TCS}$ is a $(3\times 3)$ rotation matrix.
With the positions of each of the spacecraft known in the RIC coordinate system, one can determine the additional angular velocity $\vec\omega_{\tt TCS}=\vec R_{\tt TCS}^{\rm T} \dot {\vec R}_{\tt TCS}$, although the result may be a bit complicated analytically and also less precise as it will depend on DSN tracking. Based on this, we consider $\vec\omega_{\tt TCS}$ to be known and it may be computed by following the rules used to define the TCS, as discussed in Sec.~\ref{sec:loc-tetra}. As we shall see in Sec.~\ref{sec:sagnac}, such a measurement will be provided by the Sagnac interferometers on board of each of the spacecraft.

Therefore, to describe the local spacecraft dynamics in the orbital (non-inertial) TCS reference frame, we need to add to $\vec\omega_{\tt RIC}$ from (\ref{eq:eq-om0})  additional contribution from $\vec\omega_{\tt TCS}$, namely:
  {}
\begin{equation}
\vec\omega_{\tt RIC} \qquad \rightarrow \qquad
\vec\omega=
\vec\omega_{\tt RIC}+\vec\omega_{\tt TCS},
\label{eq:angl-vel}
\end{equation}
where  $\vec\omega$ is the effective angular velocity vector of the TCS with respect to the BCRS inertial coordinate frame. 

Note that $\vec\omega_{\tt RIC}$ may be determined from navigational data and will be provided by DSN tracking. The quantity $\vec\omega_{\tt TCS}$ may be either computed using navigational data or monitored using star-trackers onboard each of the spacecraft. As we shall see in Sec.~\ref{sec:sensit}, the precision with which these two quantities may be determined individually is much worse than needed for the experiment. However, the sum of these two angular frequencies, $\vec\omega$, may be determined with a much higher accuracy using  Sagnac observables, as discussed in Sec.~\ref{sec:sagnac}, which is exactly what we need.  

By the use of kinematics, the relative acceleration
observed in the inertial reference frame $\ddot {\vec r}^{\tt }_j$ from (\ref{eq:eq-bc2rs}) and that observed locally $ \ddot{\vec r}^*_j$ in the orbital frame can be related to the measurements in the orbiting TCS reference frame as
{}
\begin{equation}
\ddot {\vec r}^{\tt }_j= \ddot{\vec r}^*_j+
2[{\vec \omega}\times \dot{\vec r}^*_j]+[{\vec \omega}\times[{\vec \omega}\times \vec r_j]]+[\dot{\vec \omega}_{\rm }\times \vec r_j],
\label{eq:eq-orb}
\end{equation}
where where the angular velocity and acceleration of the reference center ${\vec \omega}$ and $\dot{\vec \omega}$ correspond to the angular velocity and acceleration of the orbiting TCS frame with respect to BCRS frame and ${\vec \omega}$ is given by (\ref{eq:angl-vel}). Also, velocities and accelerations in the local TCS frame are given as usual:
$\dot{\vec r}^*_j=d{\vec r}^*_j/dt$ and $\ddot{\vec r}^*_j=d^2{\vec r}^*_j/dt^2$. 

As a result, combining (\ref{eq:eq-bc2rs}) and (\ref{eq:eq-orb}), we obtain the following equations of motion of a satellite $j=1,2,3$ in the TCS coordinate system associated with spacecraft \#4 but now it accounts for additional rotation from (\ref{eq:angl-vel}):
{}
\begin{equation}
\ddot{\vec r}^{*}_j=
\vec T_{\rm c}\cdot \vec r_j +\vec f_j -[{\vec \omega}\times[{\vec \omega}\times \vec r_j]]-[\dot{\vec \omega}\times \vec r_j]-
2[{\vec \omega}\times \dot{\vec r}^{*}_j].
\label{eq:eq-ocs2-n}
\end{equation}
Thus, in addition to the first two terms containing inertial accelerations from (\ref{eq:eq-bc2rs}), this equation has three other terms due to the non-inertial forces acting on the spacecraft in the TCS:  the third term is due to the centrifugal force, the forth term is due to the Euler force, and the last term is due to the Coriolis force.

For a generic angular velocity vector ${\vec \omega}=(\omega_x,\omega_y,\omega_z)$ it is convenient  to introduce the matrices ${\vec\Omega}^2$ and $\dot{\vec \Omega}$ given as
 {}
 \begin{align}
{\vec \Omega}=
 \begin{bmatrix}
0 & -\omega_z & \omega_y\\
\omega_z & 0& -\omega_x\\
-\omega_y & \omega_x & 0\\
 \end{bmatrix},
   \qquad
{\vec \Omega}^2=
 \begin{bmatrix}
-(\omega^2_y+\omega^2_z) & \omega_x \omega_y &
\omega_x \omega_z\\
\omega_x \omega_y & -(\omega^2_x+\omega^2_z) &
\omega_y \omega_z\\
\omega_x \omega_z & \omega_y \omega_z &
-(\omega^2_x+\omega^2_y) \\
 \end{bmatrix},
 \qquad
 \dot{\vec \Omega}=
 \begin{bmatrix}
0 & -\dot\omega_z & \dot\omega_y\\
\dot\omega_z & 0& -\dot\omega_x\\
-\dot\omega_y & \dot\omega_x & 0\\
 \end{bmatrix}.
 \label{eq:matOm22}
  \end{align}
These matrices allow us to simplify the cross products in (\ref{eq:eq-ocs2-n}) as
 {}
\begin{equation}
[{\vec \omega}\times[{\vec \omega}\times \vec r_j]]=
({\vec \Omega}^2 \cdot  \vec r_j),
\qquad
[\dot{\vec \omega}\times \vec r_j]=
(\dot{\vec \Omega}\cdot  \vec r_j),
\qquad
[{\vec \omega}\times \dot{\vec r}^{*}_j]=
({\vec \Omega}\cdot \dot{\vec r}^{*}_j).
\label{eq:eq-oc42}
\end{equation}
With these properties and remembering (\ref{eq:eq-ocsisws}), we present (\ref{eq:eq-ocs2-n}) as
 {}
\begin{equation}
\ddot{\vec r}^{*}_j=\Big\{\vec T_{\rm c}-{\vec \Omega}^2- \dot{\vec \Omega}\Big\}\vec r_j-
2({\vec \Omega}\cdot \dot{\vec r}^{*}_j)+\vec f_j.
\label{eq:eq-ocs3}
\end{equation}

As a result, in the TCS frame, the two spacecraft would experience different accelerations  given as
 {}
\begin{equation}
\ddot{\vec r}^{*}_{ij}=\Big\{\vec T_{\rm c}-{\vec \Omega}^2- \dot{\vec \Omega}\Big\}\vec r_{ij}-
2({\vec \Omega}\cdot \dot{\vec r}^{*}_{ij})+\vec f_{ij},
\label{eq:eq-ocsija}
\end{equation}
where we defined $\vec r_{ij}=\vec r_{j}-\vec r_{i},$ 
$ \dot{\vec r}^{*}_{ij}= \dot{\vec r}^{*}_{j}- \dot{\vec r}^{*}_{i},$
$\ddot{\vec r}^{*}_{ij}=\ddot{\vec r}^{*}_{j}-\ddot{\vec r}^{*}_{i},$ and $
\vec f_{ij}=\vec f_{j}-\vec f_{i}.$

Eq.~(\ref{eq:eq-ocsija}) may be used to recover the GGT, $\vec T_{\rm c}$ (similarly to the approach used to develop the Gravity Field and Steady-State Ocean Circulation Explorer (GOCE) mission \cite{Rummel:2011,Frommknecht-etal:2011,Siemes:2018,Sun-etal:2016}). This may be done by measuring  the relative accelerations and velocities of the spacecraft with respect to the orbital TCS frame,  $\ddot{\vec r}^{*}_{ij}$ and $ \dot{\vec r}^{*}_{ij} $, respectively, and by measuring the separation between them $\vec r_{ij}$.
For that, we could use the data provided by the laser ranging to measure $r_{ij}=|\vec r_{ij}|$. Then, by using (\ref{eq:r_coord1})--(\ref{eq:r_coord3})  with (\ref{eq:angles1})--(\ref{eq:vol-tetra}), one could take the time derivatives of the inter-spacecraft ranging data to determine  $\dot{\vec r}_{ij}$ and $\ddot{\vec r}_{ij}$.  Also, if the angular velocity $\vec \Omega(t^*)$ is known, one could use (\ref{eq:eq-ocsijs}) to determine  $\vec T_{\rm c}$.

\subsection{Solution for the gravity gradient tensor $\vec T_{\rm }$}

It is convenient to present (\ref{eq:eq-ocsija}) in the following compact matrix form
 {}
\begin{equation}
{\vec u}^{\rm }_{ij}= {\vec M} \, \vec r_{ij},
\label{eq:eq-ocsijs}
\end{equation}
where we introduced
 {}
\begin{equation}
\vec u_{ij}\equiv \ddot{\vec r}^{*}_{ij}+
2({\vec \Omega}\cdot \dot{\vec r}^{*}_{ij})-\vec f_{ij},
\qquad
{\vec M}= \vec T-{\vec \Omega}^2- \dot{\vec \Omega}.
\label{eq:eq-ocsi3def}
\end{equation}

Clearly, given the fact that we are dealing with a tetrahedron that has six edges, there are several ways to develop an estimate for $\vec M$ by using (\ref{eq:eq-ocsijs}).  Indeed, one either i) forms a $3\times 6$ matrix equation by using all the six inter-satellite range measurements, $\vec r_{ij}$, available with the tetrahedron, ii) may use three edges sharing the same vertex to from a $3\times3$ matrix equation to determine $\vec M$ at a particular vertex, or iii) may do the same procedure as just mentioned in item ii) for all four vertices within the tetrahedron  to develop a weighted sum needed to provide a more realistic system description.  

Clearly, all the approaches mentioned above must be explored for the ultimate mission development.  However, for our purposes aimed at a feasibility study and a preliminary error budget development, it is sufficient to use the approach identified in item ii) above. For that, we assume that the origin of our reference frame is associated with one of the spacecraft, namely spacecraft \#4. This assumption, will allow us to develop the relevant solution for $\vec M$. Therefore, considering that the origin of the TCS is placed at spacecraft \#4,  we may write (\ref{eq:eq-ocsijs}) in the following matrix form:
 {}
\begin{align}
 \bf U = \bf M \bf R,
 \label{eq:mat1234}
  \end{align}
where matrices $\bf U$, $\bf M$, and $\bf R$ are given as
    \begin{align}
\bf U= \begin{bmatrix}
u_{41}^x & u_{42}^x & u_{43}^x\\
u_{41}^y & u_{42}^y & u_{43}^y \\
u_{41}^z & u_{42}^z & u_{43}^z\\
 \end{bmatrix},
 \qquad
 \bf M=
  \begin{bmatrix}
M_{11} & M_{12} & M_{13}\\
M_{21} & M_{22} & M_{23}\\
M_{31} & M_{32} & M_{33}\\
 \end{bmatrix},
  \qquad
 \bf R=
 \begin{bmatrix}
r_{41}^x & r_{42}^x & r_{43}^x\\
r_{41}^y & r_{42}^y & r_{43}^y \\
r_{41}^z & r_{42}^z & r_{43}^z\\
 \end{bmatrix},
 \label{eq:mat123d}
  \end{align}
where we rely only on the ranges and accelerations associated with the edges sharing the same vertex \#4. Clearly, similar expressions may be developed for any of the vertices within the tetrahedron. 

Assuming that  $\det({\vec R})=(\vec r_{41}\cdot[\vec r_{42}\times \vec r_{43}])\not =0$, the inverse of $ \bf R$ has the form:
{}
\begin{align}
{ \bf R}^{-1}=\frac{1}{\det(\bf R)}
 \begin{bmatrix}
[\vec r_{42}\times \vec r_{43}]^{\rm T}\\[2pt]
[\vec r_{43}\times \vec r_{41}]^{\rm T} \\[2pt]
[\vec r_{41}\times \vec r_{42}]^{\rm T}\\[2pt]
 \end{bmatrix}=
\Big(\frac{\tilde{\vec n}_{41}}{r_{41}},\frac{\tilde{\vec n}_{42}}{r_{42}},\frac{\tilde{\vec n}_{43}}{r_{43}}\Big),
 \label{eq:mat123f}
  \end{align}
where subscript $(...)^{\rm T}$ indicates a transposed quantity and we introduced the unit vectors of the reciprocal coordinate basis \cite{Wheeler-1998}
  {}
\begin{eqnarray}
\tilde {\vec n}_{41}&=&
\frac{[\vec n_{42}\times \vec n_{43}]}{({\vec n}_{41}\cdot [\vec n_{42}\times \vec n_{43}])}, \qquad
\tilde {\vec n}_{42}=
\frac{ [\vec n_{43}\times \vec n_{41}]}{({\vec n}_{42}\cdot [\vec n_{43}\times \vec n_{41}])},
\qquad
\tilde {\vec n}_{43}=
\frac{[\vec n_{41}\times \vec n_{42}]}{({\vec n}_{43} \cdot [\vec n_{41}\times \vec n_{42}])}.
\label{eq:recipr}
  \end{eqnarray}
  
We note that, in the case of a right tetrahedron, the unit vectors (\ref{eq:recipr}) would represent the base vectors of the orthogonal coordinate system.  However, our tetrahedron is rather generic, so these vectors account for non-orthogonality of the vectors emanating from vertex \#4. Clearly, when the volume of the tetrahedron collapses, the denominator in (\ref{eq:recipr}) vanishes. In this case one looses the ability to determine the trace of the GGT (\ref{eq:mat12=2}). 

As a result, the solution for $\bf M$ is given as
\begin{align}
{\bf M}=  \bf U  { \bf R}^{-1}.
\label{eq:mat123f+}
\end{align}

For studying the Laplace equation, $\nabla^2 U=0$, we do not need the entire expression for $\bf M$, but only its trace, that with (\ref{eq:recipr}) takes a rather convenient and informative form
  {}
 \begin{align}
{\rm tr} ({\bf M})_4= \frac{({\vec u}_{41}\cdot \tilde{\vec n}_{41})}{r_{41}}+
\frac{({\vec u}_{42}\cdot \tilde {\vec n}_{42})}{r_{42}}
+\frac{({\vec u}_{43} \cdot \tilde {\vec n}_{43})}{r_{43}},
 \label{eq:mat12=2}
  \end{align}
where the accelerations in the orbital TCS frame are given by (\ref{eq:eq-ocsi3def}).

As it stands, the result (\ref{eq:mat12=2}) with the base vectors $\tilde {\vec n}_{41}, \tilde {\vec n}_{42}, \tilde {\vec n}_{43}$  (\ref{eq:recipr}), is the solution for the trace of the GGT describing the most general situation with a generic and elastically-evolving tetrahedron. We observe that, in the case of a right tetrahedron, expressions (\ref{eq:recipr}) become the base vectors of the ortho-normal coordinate system with the origin at vertex \#4: $\tilde {\vec n}_{41} \rightarrow {\vec e}_{41}=(1,0,0)$, $\tilde {\vec n}_{42} \rightarrow {\vec e}_{42}=(0,1,0)$, and $\tilde {\vec n}_{43} \rightarrow {\vec e}_{43}=(0,0,1)$. As a result, in this case, (\ref{eq:mat12=2}) would result in a sum of the acceleration projections on the three base vectors divided by the lengths of the corresponding edges. 

In essence,  result (\ref{eq:mat12=2}) constitutes a finite difference version of the trace of the GGT (i.e., a finite difference version of the Laplace equation presented in a general non-orthogonal coordinate system, providing a generalization for the basic equation used in gravitational gradiometry \cite{Rummel:2011} to determine ${\rm tr} ({\bf M})$ -- our primary quantity of interest. Clearly, the shorter the separation between the spacecraft, the more this quantity would resemble its infinitesimal version.

To establish the explicit dependence of ${\rm tr} ({\bf M})_4$ on the observable quantities, we remember the definitions for ${\vec\Omega}^2$ and $\dot{\vec \Omega}$  from (\ref{eq:matOm22}) that yield the following identities
 {}
 \begin{align}
{\rm tr} ({\bf  \Omega}^2)=-2(\omega^2_x+\omega^2_y+\omega^2_z)=-2\omega^2, \qquad
{\rm tr} (\dot{\vec \Omega})=0,
 \label{eq:ome}
  \end{align}
where  $ \omega=|\vec \omega|$ is from (\ref{eq:angl-vel}). Note that the second result in (\ref{eq:ome}), shows that the Euler force does to not contribute to ${\rm tr} ({\bf M})_4$. As a result, remembering that the definition for ${\vec M}$ from (\ref{eq:eq-ocsi3def}) and using (\ref{eq:ome}), we have
 {}
\begin{align}
{\rm tr} ({\bf M})_4=
{\rm tr} ({\vec T})_4+2\omega^2,
 \label{eq:dgd}
\end{align}
that represents the Laplacian of the effective potential energy for an orbiting platform, $E=GM/r+{\textstyle\frac{1}{2}}L^2/m^2r^2$, where $\vec L=m{\vec \omega} r^2$ is the angular momentum with $\omega$ being the angular velocity \cite{Landau-Lifshitz:1976}. So, the first term in (\ref{eq:dgd}) is due to the trace of the inertial GGT and the second one is due to the gradients of the centrifugal potential.      
  
 Next, collecting the results (\ref{eq:dgd}) and $\vec u_{4i}$ from  (\ref{eq:eq-ocsi3def}), and remembering $({\vec \Omega}\cdot \dot{\vec r}^{*}_j)=[{\vec \omega}\times \dot{\vec r}^{*}_j]$, 
in the case, when the tetrahedron's volume is not vanishing (i.e., when $({\vec n}_{41}\cdot [\vec n_{42}\times \vec n_{43}])\not=0$, meaning that in (\ref{eq:recipr}) the denominator is non-vanishing), we can present (\ref{eq:mat12=2}) in the following form
{}
\begin{eqnarray}
{\rm tr} ({\bf T})_4&+&2\omega^2=\nonumber\\
&&\hskip -30 pt =\, 
\frac{\Big(\big(\ddot{\vec r}^{*}_{41}+2[{\vec \omega}\times \dot{\vec r}^{*}_{41}]-{\vec f}_{41}\big)\cdot \tilde{\vec n}_{41}\Big)}{r_{41}}
+
\frac{\Big(\big(\ddot{\vec r}^{*}_{42}+2[{\vec \omega}\times \dot{\vec r}^{*}_{42}]-{\vec f}_{42}\big)\cdot \tilde{\vec n}_{42}\Big)}{r_{42}}
+
\frac{\Big(\big(\ddot{\vec r}^{*}_{43}+[{\vec \omega}\times \dot{\vec r}^{*}_{43}]-{\vec f}_{43}\big) \cdot \tilde{\vec n}_{43}\Big)}{r_{43}},~~~
 \label{eq:dgd+3}
  \end{eqnarray} 
 which shows the relationship wherein the trace of the tensor ${\rm tr} ({\bf T})_4$ explicitly depends on observable quantities, denoted by $\vec r_{ij}$ and their respective time derivatives in the TCS orbital frame, forces ${\vec f}_{ij}$ (\ref{eq:eq-ff}) and angular frequency $\vec \omega$ (\ref{eq:angl-vel}). The quantities $\vec r_{ij}$  will be  measured by laser metrology, so as the  unit vectors of the reciprocal coordinate basis (\ref{eq:recipr}). The vector ${\vec f}_{ij}$ represents forces that will be measured directly measured by the atom interferometry and calibrated out based on their predefined analytical structure, as shown by (\ref{eq:eq-ff}).  The angular velocity $\vec \omega$ will be provided by the Sagnac interferometry, as discussed  Sec.~\ref{sec:sagnac}.
 
Expression (\ref{eq:dgd+3}) describes the trace of the GGT in the orbital coordinate frame. It provides a very sensitive test for any anomalous dynamical field that may be present in the solar system. This expression accounts for the fact that the orbital frame is non-inertial that is evidenced by the presence of  terms with the angular velocity $\vec\omega$ -- the terms due to centrifugal and Coriolis forces.  As we shall see in Sec.~\ref{sec:sensit}, all the quantities relevant to the tetrahedron itself -- intersatellite ranges, $r_{ij}$, will be provided by the laser ranging, and range-rates, $\dot r_{ij}$, and range accelerations, $\ddot r_{ij}$,  will be computed based on the range data. The only quantity that is not provided by the local measurements, is the angular frequency. If this quantity is known to a sufficient precision, then ${\rm tr} ({\bf T})_4$ can be computed. As we shall see in Sec.~\ref{sec:sagnac}, the angular frequency, $\vec \omega$, will be provided by the Sagnac interferometry.

Based on our analysis (see Sec.~\ref{sec:eccentric-orb}), we know that to set up a tetrahedral formation with desirable properties, the orbits of the spacecraft, ${\vec R}_j$, should have the same eccentricity and semi-major axes. The other Keplerian orbital elements are different and are chosen in a such a way so that the choice would allow to maximize the volume of the resulting tetrahedron, at least for some parts of the orbit. Also, as the spacecraft move in their heliocentric orbits, the  inter-spacecraft separations vary significantly causing the tetrahedral structure to evolve, while the tetrahedron itself undergoes a complex rotation in the orbital frame. Periodically, when one of the spacecraft crosses the instantaneous plane formed by the other three spacecraft, the volume of the tetrahedron collapses. Expression (\ref{eq:dgd+3}) shows how such a complex dynamics on the orbital frame affects the determination of ${\rm tr} ({\bf T})_4$.

\subsection{Required measurement sensitivities}
\label{sec:sensit}

It is instructive to present expression (\ref{eq:dgd+3}) in terms of the TCS coordinates introduced in Sec.~\ref{sec:loc-tetra}. Clearly, as ${\rm tr} ({\bf T})_4$ is a scalar, this result will be the same in any other coordinate system.  However, the chosen TCS coordinate system allows us to express all the quantities involved in terms of the six tetrahedral edges measured by laser ranging. First of all, using (\ref{eq:r_coord1})--(\ref{eq:r_coord3})  and (\ref{eq:angles1})--(\ref{eq:angles3}), we establish the structure of some of the terms in (\ref{eq:dgd+3}), namely:
 {}
\begin{align}
[\vec n_{42}\times \vec n_{43}] &=\Big(\sin\alpha_{12}\sin\beta_3,\, -\cos\alpha_{12}\sin\beta_3,\,
\cos\alpha_{12}\sqrt{\sin^2\alpha_{13}-\sin^2\beta_3}-\sin\alpha_{12}\cos\alpha_{13}\Big),\\
[\vec n_{43}\times \vec n_{41}] &=\Big(0,\, \sin\beta_3,\, -\sqrt{\sin^2\alpha_{13}-\sin^2\beta_3}\Big),\\
 [\vec n_{41}\times \vec n_{42}] &=\Big(0,\, 0,\,\sin\alpha_{12}\Big),
 \qquad\qquad\qquad\qquad
({\vec n}_{41}\cdot [\vec n_{42}\times \vec n_{43}])=
\sin\alpha_{12}\sin\beta_3,
 \label{eq:exp2}
  \end{align}
  which allow us to compute the unit vectors of the reciprocal coordinate basis $\tilde {\vec n}_{41}, \tilde {\vec n}_{42}, \tilde {\vec n}_{43}$ from (\ref{eq:recipr}), expressing them via the quantities  $\alpha_{12}, \alpha_{13}$, and $\beta_{3}$ given by (\ref{eq:angles1})--(\ref{eq:vol-tetra}) that are observed with  laser interferometric ranging.

Now, treating $\ddot{\vec r}_{4i}$ as the second time derivative of ${\vec r}_{4i}$ that is given by (\ref{eq:r_coord1})--(\ref{eq:r_coord3}) together with  (\ref{eq:angles1})--(\ref{eq:angles3}), we have the following result for acceleration-dependent contributions (i.e., $\propto \ddot{\vec r}_{4i}$)  to  ${\rm tr}(\vec T)_4$:
 {}
\begin{eqnarray}
{\rm tr} ({\bf T})^{\rm accel}_4&=&
\frac{\ddot{r}_{41}}{r_{41}}+
\frac{\ddot{r}_{42}}{r_{42}}+\frac{\ddot{r}_{43}}{r_{43}}
+\frac{(\sin\alpha_{12})^{..}}{\sin\alpha_{12}}
+\frac{(\sin \beta_3)^{..}}{\sin\beta_3}+
\frac{2\dot{r}_{42}}{r_{42}}\frac{(\sin\alpha_{12})^.}{\sin\alpha_{12}}
+\frac{2\dot{r}_{43}}{r_{43}}\frac{(\sin \beta_3)^.}{\sin\beta_3}.
 \label{eq:dg2fg}
  \end{eqnarray}
Note that  ${\rm tr}(\vec T)_4$  is independent of the signs of $\sin\alpha_{12}$ and $\sin\beta_3$, as it should be. 

Considering  this acceleration-dependent contribution (\ref{eq:dg2fg}), we see that it has the following generic structure:
 {}
\begin{eqnarray}
{\rm tr} ({\bf T})^{\rm accel}_4&\simeq&
\sum_{i=1,2,3}\Big\{\frac{\ddot{r}_{4i}}{r_{4i}}+
\Big(\frac{\dot{r}_{4i}}{r_{4i}}\Big)^2\Big\},
 \label{eq:dg2fg=}
  \end{eqnarray}
that implies that in order to determine ${\rm tr} ({\bf T})_4$ with a single measurement accuracy of $\delta ({\rm tr} ({\bf T})_4)\sim 10^{-21}~{\rm s}^{-2}$,  range-rates must be available with the accuracy of $\delta \dot{r}_{4i} \lesssim r_{4i} \sqrt{\delta{\rm tr} ({\bf T})_4}\sim 3.2\times 10^{-5}~{\rm m/s}$ and the line-of-sight accelerations must be known with the accuracy better than $\delta \ddot{r}_{4i} \lesssim r_{4i} \delta{\rm tr} ({\bf T})_4\sim 1\times 10^{-15}~{\rm m/s}^{2}$, thus setting the requirements on the accuracy of the laser ranging measurements. These requirements are summarized in Table~\ref{tb:require-params}. Assuming that the measurement errors are uncorrelated and normally distributed, and the ranging instruments allow achieving $\delta ({\rm tr} ({\bf T})_4)\sim 10^{-21}~{\rm s}^{-2}$ in 10~s, the mission precision of $\delta ({\rm tr} ({\bf T})_4)\sim 10^{-24}~{\rm s}^{-2}$ will be reached in $\sim 4$ months.

Next, we compute the $\vec f_{4i}$-dependent contributions  to  ${\rm tr}(\vec T)_4$ in (\ref{eq:dgd+3}):
 {}
\begin{eqnarray}
{\rm tr} ({\bf T})_4^{\rm forces}&=&
-\frac{1}{r_{41}}\Big\{f_{41x}-f_{41y}\cot \alpha_{12}+\frac{f_{41z}}{\sin\beta_3}\Big(\cot \alpha_{12}\sqrt{\sin^2\alpha_{13}-\sin^2\beta_3}-\cos\alpha_{13}\Big)\Big\}-\nonumber\\
&&-\,
\frac{1}{r_{42}\sin\alpha_{12}}\Big\{f_{42y}-\frac{f_{42z}}{\sin\beta_3}\sqrt{\sin^2\alpha_{13}-\sin^2\beta_3}\Big\}-\frac{f_{43z}}{r_{43}\sin\beta_{3}}.
 \label{eq:dg2fg-f}
  \end{eqnarray}

Similarly to (\ref{eq:dg2fg=}), we see that the $\vec f_{4i}$-dependent contribution to (\ref{eq:dg2fg-f}) behaves as
 {}
\begin{eqnarray}
{\rm tr} ({\bf T})_4^{\rm forces}&\simeq&
\sum_{i=1,2,3} \frac{f_{4i}}{r_{4i}},
 \label{eq:dg2fg-f=}
  \end{eqnarray}
  which implies that the non-gravitational forces must be compensated to below $\delta f_{4i} \lesssim r_{4i} \delta{\rm tr} ({\bf T})_4\sim 1\times 10^{-15}~{\rm m/s}^{2}$, which may be achieved by using atom interferometers.
  
We evaluate the Coriolis terms due to rotation of the TCS with respect to the BCRS to be:
  {}
\begin{eqnarray}
{\rm tr} ({\bf T})_4^{\rm Cor}&=&2\bigg\{
\frac{\omega_x}{\sin\beta_3}\Big\{\sqrt{\sin^2\alpha_{13}-\sin^2\beta_3}\Big(\frac{\dot r_{43}}{r_{43}}-\frac{\big(r_{42}\sin\alpha_{12}\big)^.}{r_{42}\sin\alpha_{12}}\Big)+\Big(\sqrt{\sin^2\alpha_{13}-\sin^2\beta_3}\Big)^.\Big\}+\nonumber\\
&&\hskip-9pt +\,
\frac{\omega_y}{\sin\beta_3}\Big\{\cos\alpha_{13}\Big(\frac{\dot r_{41}}{r_{41}}-\frac{\dot r_{43}}{r_{43}}\Big)+\sqrt{\sin^2\alpha_{13}-\sin^2\beta_3}\Big(\frac{\big(r_{42}\cos\alpha_{12}\big)^.}{r_{42}\sin\alpha_{12}}-\cot\alpha_{12}\frac{\dot r_{41}}{r_{41}}\Big)
-\big(\cos\alpha_{13}\big)^.\Big\}+\nonumber\\
&&\hskip 7pt +\,
\omega_z\Big\{\frac{\big(r_{42}\cos\alpha_{12}\big)^.}{r_{42}\sin\alpha_{12}}-\cot\alpha_{12}\frac{\dot r_{41}}{r_{41}}\Big\}\bigg\}.
 \label{eq:dg2fg-om}
 \end{eqnarray}
 
With the help of (\ref{eq:dg2fg-om}), we observe that the largest $\vec\omega$-dependent terms exhibit the following generic behavior
  {}
\begin{eqnarray}
{\rm tr} ({\bf T})_4^{\rm Cor}&\simeq&-
2\sum_{i,j,k}\omega_i\Big(\frac{\dot r_{4j}}{r_{4j}}-\frac{\dot r_{4k}}{r_{4k}}\Big).
 \label{eq:dg2fg-om=}
  \end{eqnarray}

 Assuming that the nominal range-rates of $\dot r_{4i}\sim 0.20~{\rm m/s}$ (see (\ref{eq:vel-acc})), especially at the apogee, (\ref{eq:dg2fg-om=}) implies that the angular velocity must be known with accuracy of $\delta \omega \lesssim (r_{4i}/\dot r_{4i})\textstyle{\frac{1}{2}} \delta{\rm tr} ({\bf T})_4\simeq 2.5\times 10^{-15}~{\rm s}^{-1}$.  This is a challenging requirement\footnote{Potentially,  long-baseline AI can be used to provide the needed precision in measuring such a rotation.} that eliminates the use of both the DSN tracking and on-board star-trackers as the means to determine $\vec \omega(t)$. (A similar point on the importance of Coriolis forces in the rotational sensitivity of atomic interferometers was discussed in \cite{Lan-etal:2012}.)  As will be discussed in Sec.~\ref{sec:sagnac}, Sagnac observables are capable of providing the needed precision. 
 
 Finally, we evaluate the centrifugal terms due to rotation of the TCS with respect to the BCRS as below:
  {}
\begin{eqnarray}
{\rm tr} ({\bf T})_4^{\rm centrif}&=&-2\omega^2.
 \label{eq:dg2fg-om+ce}
 \end{eqnarray}

One can see that this term yields requirement on the precision of $\omega$: $\delta \omega_{\tt CF} \lesssim ({\textstyle{\frac{1}{2}}\delta{\rm tr} ({\bf T})_4})^\frac{1}{2}\simeq 2.24\times 10^{-11}~{\rm s}^{-1}$, which is less stringent compared to that derived from (\ref{eq:dg2fg-om=}) presented by the need to compensate for the Coriolis forces.

Results (\ref{eq:dg2fg})--(\ref{eq:dg2fg-om+ce}) express the different terms in the expression for ${\rm tr}(\vec T)$ in (\ref{eq:dgd+3}).   We can use these expressions to evaluate the  accuracy needed to conduct the experiment. For that, we assume that spacecraft \#4, the origin of our TCS, moves along the heliocentric orbit with semi-major axis  and eccentricity of $a\simeq 1$~AU and $e\simeq 0.6$, correspondingly. Also, we assume that the nominal separation between the spacecraft is $r_{ij}\simeq 10^3$~km.

\begin{table*}[t!]
\vskip-15pt
\caption{Select set of the GDEM mission requirements, along with corresponding symbols used in the text.
\label{tb:require-params}}
\begin{tabular}{|l|c|c|}\hline
Parameter  &Symbol  &Value\\\hline\hline
Intersatellite range-rate &$\delta \dot{r}_{4i}$ &\phantom{0}$3\times 10^{-5}~{\rm m/s}$\phantom{/s}\\
Line-of-sight acceleration &$\delta \ddot{r}_{4i}$ &\phantom{00}$1\times 10^{-15}~{\rm m/s}^{2}$\phantom{/s}\\
Non-gravitational forces  &$\delta f_{4i}$ &\phantom{0}$1\times 10^{-15}~{\rm m/s}^{2}$\phantom{/s}\\
Angular velocity  &$\delta \omega$&\phantom{0.}$2\times 10^{-15}~{\rm s}^{-1}$\phantom{k}\\
Centrifugal acceleration & $\delta \omega_{\tt CF}$ & \phantom{0}$2\times 10^{-11}~{\rm s}^{-1}$\phantom{k}\\\hline
\end{tabular}
\end{table*}
\vskip-0pt

As a result, we are able to express all the quantities involved in the determination of the trace of ${\rm tr} ({\bf T})_4$ via two types of observables, namely: i) laser ranging measurements that will provide time series of the inter-satellite ranges, $r_{ij}$, that can be time-differentiated to derive range-rate and line of sight accelerations, $\dot r_{ij}$, $\ddot r_{ij}$; and ii) the angular velocity of the non-inertial TCS, $\vec \omega$, with respect to inertial coordinates of the BCRS. Therefore,  the knowledge of $\vec \omega$ is critical for the success of the experiment. Below, we consider two methods to determine the angular velocity of the TCS with respect to the BCRS, including the kinematic determination and that relying on the Sagnac interferometry.

\subsection{Kinematic equation to determine $\dot{\vec \Omega}$}

One way to measure $\vec \omega$ is to use the same set of equations and related observables that was used to derive ${\rm tr} ({\bf T})_4$ given by (\ref{eq:dgd+3}), as is typically done by gravity gradiometry missions \cite{Rummel:2011,Sun:2016}. For that, using the definition for $ {\bf M}$ from (\ref{eq:eq-ocsi3def}), and relying on the fact that GGT is a symmetric matrix, $\vec T^{\rm T}_{\rm c} = \vec T_{\rm c}$, and also $(\vec \Omega^2)^{\rm T}= \vec \Omega^2$, $\dot{\vec \Omega}^{\rm T}=-\dot{\vec \Omega}$,  we have:
{}
 \begin{align}
\Big(\vec M^{\rm T}-\vec M\Big)=\Big(\vec T_{\rm c} -\vec \Omega^2- \dot{\vec \Omega}\Big)^{\rm T}-\Big(\vec T_{\rm c} -\vec \Omega^2- \dot{\vec \Omega}\Big)=2\dot{\vec \Omega},\qquad \Rightarrow \qquad \dot{\vec \Omega}=
{\textstyle\frac{1}{2}}\Big(\vec M^{\rm T}-\vec M\Big).
 \label{eq:mat123dd}
  \end{align}
Then, with the help of (\ref{eq:mat123f+}), the solution to $\dot{\vec \Omega}$ may be given as follows
 {}
\begin{align}
\dot{\vec \Omega}&=
\begin{bmatrix}
&0,&-\dot\omega_z,&\dot\omega_y\\[0pt]
&\dot\omega_z,&0,&-\dot\omega_x
\\[0pt]
&-\dot\omega_y, &\dot\omega_x,&0\\
 \end{bmatrix},
 \label{eq:om-dot-co}
  \end{align}
where $\dot{\vec \omega}=(\dot\omega_x, \dot\omega_y, \dot\omega_z)$ is given as
{}
\begin{align}
\dot{\vec \omega}&=\frac{[\vec u_{41}\times \tilde {\vec n}_{41} ]}{2r_{41}}+
\frac{[\vec u_{42}\times  \tilde{\vec n}_{42}]}{2r_{42}}+
\frac{[\vec u_{43}\times \tilde{\vec n}_{43}]}{2r_{43}},
 \label{eq:om-vec}
\end{align}
where $\tilde {\vec n}_{41},\tilde {\vec n}_{42},\tilde {\vec n}_{43}$ are the unit vectors of the reciprocal basis given by (\ref{eq:recipr}).
It is interesting to note that the structure of (\ref{eq:om-vec}) is similar to that of the GGT trace ${\rm tr}(\vec T_4)$ from (\ref{eq:mat12=2}). However, as opposed to the scalar product of $\vec u_{ij}/r_{ij}$ to the vectors of the reciprocal basis present in the result for ${\rm tr}(\vec T_4)$,  $\dot{\vec \omega}$ has the cross product.

Next, substituting in (\ref{eq:om-vec}), the expression for $\vec u_{ij}$ from (\ref{eq:eq-ocsi3def}), we see that the temporal evolution of the angular velocity vector is driven by the Euler and Coriolis forces and is governed by the following differential equation: 
{}
\begin{align}
\dot{\vec \omega}&=\frac{\big[\big(\ddot{\vec r}^{*}_{41}+2[{\vec \omega}\times \dot{\vec r}^{*}_{41}]-{\vec f}_{41}\big)\times \tilde{\vec n}_{41}\big]}{2r_{41}}+
\frac{\big[\big(\ddot{\vec r}^{*}_{42}+2[{\vec \omega}\times \dot{\vec r}^{*}_{42}]-{\vec f}_{42}\big)\times \tilde{\vec n}_{42}\big]}{2r_{42}}+
\frac{\big[\big(\ddot{\vec r}^{*}_{43}+2[{\vec \omega}\times \dot{\vec r}^{*}_{43}]-{\vec f}_{43}\big)\times \tilde{\vec n}_{43}\big]}{2r_{43}}.
 \label{eq:om-vec5}
  \end{align}
  
Eq.~(\ref{eq:om-vec5}) is the differential equation that establishes the temporal evolution of  ${\vec \omega}(t)$. 
In principle, if precise knowledge of the relevant initial and/or boundary conditions is available, one can solve this equation to determine ${\vec \omega}(t)$ \cite{Siemes-2012}.  

How can we accurately determine the value of $\omega$ for the spacecraft within the tetrahedron? One approach might be to employ DSN tracking or other external ranging methods. However, these techniques, while direct, might not achieve the desired precision. It's important to note that even with projected advancements in tracking accuracy, these methods may still fall short of the precision needed. 
Thus,  practical use of (\ref{eq:om-vec5}) may be limited, requiting other methods.  In Sec.~\ref{sec:sagnac} we  demonstrate that Sagnac observables successfully address that need.

\subsection{Generalized Sagnac observables}
\label{sec:sagnac}

Another method to determine the angular velocity ${\vec \omega}$ is to use  Sagnac observables to do the job. In fact, Sagnac interferometers are commonly used as rotation sensors. For that, one considers the difference between the optical paths of the counter-propagating beams in a rotating Sagnac interferometer  that may be modeled as \cite{Shaddock:2004}
{}
\begin{align}
\Delta L_{412}&=\ell^+-\ell^-=
\big(r_{41}+r_{12}+r_{24}\big)-
\big(r_{14}+r_{21}+r_{42}\big)=
\frac{4}{c}\big(\vec\omega\cdot \vec A_{412}\big),
 \label{eq:sint412}
  \end{align}
 where  $\vec\omega$ is the angular velocity, $\vec A_{412}$ is the oriented area enclosed by the optical paths, and $c$ is the speed of light. So, if only one triangle is available,  Sagnac signal provides information about  projection of the angular velocity vector on the oriented area of the triangle. However, in the case of a tetrahedron, one has access to three projections of the same vector on three different oriented areas, thus enabling the determination of all the three components of $\vec\omega$.

Note that  expression (\ref{eq:sint412}) does not account for the fact that all the triangles at any face of the tetrahedron evolve on the time scale of light propagation between the spacecraft.  As a result, not only all the six edges of the tetrahedron change lengths, the tetrahedron itself undergoes a kinematic rotation, so that all the oriented areas change their size and spatial orientation. The situation of using Sagnac interferometers at four vertices of an elastic tetrahedron to determine its inertial orientation is new and had not been previously considered \cite{Wang:2004,Ori-Avron:2016}. Therefore, we need to present the model that could be used to determine $\omega$ under such conditions. We do that next.

Here, we generalize expression (\ref{eq:sint412})  to account for the tetrahedron's elasticity.  To do that, we directly evaluate the optical paths in the co- and contra-rotating directions. We begin with the triangle $\{4,1,2\}$ and present the equations that describe the light propagation along the different edges in two opposite directions. For that, we rely on the fact that the origin of our TCS coordinate system is at spacecraft \#4, thus, $\vec r_4=0$ and $\dot{\vec r}_4=0$.

First, we consider the clockwise direction: The logic of this measurement is summarized as follows:
\begin{enumerate}[i).]
\item  A light signal is emitted at spacecraft \#4 at time $t_0$ and travels toward spacecraft \#1;
\item   The signal is coherently received at spacecraft \#1 at time $t_1=t_0+ \Delta t_{41}$ and retransmitted toward spacecraft \#2;
\item  The signal is coherently received at spacecraft \#2 at time $t_2=t_1+ \Delta t_{12}\equiv t_0+ \Delta t_{41}+\Delta t_{12}$ and retransmitted back to spacecraft \#4, where it is received at time $t_4=t_2+\Delta t_{24}\equiv t_0+ \Delta t_{41}+\Delta t_{12}+\Delta t_{24}$.
 \end{enumerate}
 Assuming that the individual light transit times $\Delta t_{ij}$ are small, such that $c\Delta t_{ij}\simeq r_{ij}\ll R_{\rm c}$, we can model the vectors that describe the paths that the light signal travelled while propagating alone the different sides of the triangle  as
 {}
\begin{align}
\vec r_{41}(t_1)&= \vec r_{41}(t_0)+ \dot{\vec r}_{41}(t_0)\Delta t_{41}+{\cal O}(\Delta t^2),
\label{eq:r41}\\
\vec r_{12}(t_2)=\vec r_{42}(t_2)-\vec r_{41}(t_1)&= \vec r_{12}(t_0)+ \dot{\vec r}_{12}(t_0)\Delta t_{41}+\dot{\vec r}_{42}(t_0)\Delta t_{12}+{\cal O}(\Delta t^2),\\
\vec r_{24}(t_4)\equiv \vec r_{24}(t_2)&= \vec r_{24}(t_0)+ \dot{\vec r}_{24}(t_0)\Big(\Delta t_{41}+\Delta t_{12}
\Big)+{\cal O}(\Delta t^2),
 \label{eq:cor}
\end{align}
where we relied on the fact that spacecraft \# 4  is the origin of the chosen orbital coordinate system, as detailed in  Sec.~\ref{sec:loc-tetra}. Consequently, its position vector is $\vec r_4=0$, justifying the following identity $\vec r_{24}(t_4)=\vec r_{4}(t_4)-\vec r_{2}(t_2)\equiv \vec r_{24}(t_2)$.

Note that in (\ref{eq:r41})--(\ref{eq:cor}), we consider only the terms linear with respect to $\Delta t$. As a result, the light trajectories, that,  in this approximation may be taken to be straight lines, are given as below
{}
\begin{align}
 r_{41}(t_1)&= r_{41}(t_0)+ (\vec n_{41}\cdot\dot{\vec r}_{41})\Delta t_{41}+{\cal O}(\Delta t^2),\label{eq:om41a}\\
 r_{12}(t_2)&=  r_{12}(t_0)+ (\vec n_{12}\cdot\dot{\vec r}_{12})\Delta t_{41}+(\vec n_{12}\cdot\dot{\vec r}_{42})\Delta t_{12}+{\cal O}(\Delta t^2), \label{eq:om41}\\
r_{24}(t_4)&=  r_{24}(t_0)+ (\vec n_{24}\cdot\dot{\vec r}_{24})\Big(\Delta t_{41}+\Delta t_{12}
\Big)+{\cal O}(\Delta t^2).
 \label{eq:cor2}
\end{align}

We remember that TCS is a non-inertial reference frame, which requires that the  time derivatives must include the angular velocity contributions $\vec\omega$, namely $\dot{\vec r}_{ij} \rightarrow \dot{\vec r}_{ij}+[\vec \omega\times{\vec r}_{ij}]$. It is clear that   $(\vec n_{ij}\cdot\dot{\vec r}_{ij})$  will be unchanged as  $(\vec n_{ij}\cdot\ [\vec \omega\times{\vec r}_{ij}])=0$. The only place where $\vec \omega$ does not vanish is the last term in (\ref{eq:om41}), namely
  {}
\begin{align}
(\vec n_{12}\cdot\dot{\vec r}_{42}) \rightarrow
(\vec n_{12}\cdot\dot{\vec r}_{42}) + (\vec n_{12}\cdot[\vec \omega\times{\vec r}_{42}])\equiv (\vec n_{12}\cdot\dot{\vec r}_{42}) + \frac{2}{r_{12}} (\vec \omega \cdot \vec A_{412}).
\label{eq:om412}
\end{align}

We realize that (\ref{eq:om41a})--(\ref{eq:cor2}) may be expressed via the travel times for the light to move 
along the different sides of the triangle, $c\Delta t_{ij}=r_{ij}$.  Using this fact and accounting for (\ref{eq:om412}), we present these equations below, valid to ${\cal O}(\Delta t^2)$:
 {}
\begin{align}
c\Delta t_{41}&= r_{41}+ \dot{ r}_{41}\Delta t_{41},
\label{eq:om41a2}\\
c\Delta t_{12}&=  r_{12}+ \dot{ r}_{12}\Delta t_{41}+\Big((\vec n_{12}\cdot\dot{\vec r}_{42})+\frac{2}{r_{12}} (\vec \omega \cdot \vec A_{412})\Big)\Delta t_{12},
\label{eq:om412a}\\
c\Delta t_{24}&=
r_{24}+ \dot{ r}_{24}\Big(\Delta t_{41}+\Delta t_{12}\Big),
 \label{eq:cor22}
\end{align}
where, for brevity, we used $r_{ij}\equiv r_{ij}(t_0)$  and  also $ (\vec n_{ij}\cdot\dot{\vec r}_{ij})=\dot r_{ij}(t_0)\equiv \dot r_{ij}$. These equations may be used to determine the light transit times $\Delta t_{ij}$ by developing an approximate solution with respect to a small parameter $(\dot r_{ij}/c)$:
 {}
\begin{align}
c\Delta t_{41}&
\simeq  r_{41}+c^{-1}r_{41}\dot{ r}_{41} +{\cal O}(c^{-2}),
\label{eq:om41a3}\\
c\Delta t_{12}&
\simeq r_{12}+c^{-1}\Big(r_{41}\dot{ r}_{12}+(\vec r_{12}\cdot\dot{\vec r}_{42})+2 (\vec \omega \cdot \vec A_{412})\Big)
+{\cal O}(c^{-2}),
\label{eq:om413}\\
c\Delta t_{24}&
\simeq r_{24}+ c^{-1}\dot{ r}_{24}\Big(r_{41}+r_{12}
\Big)+{\cal O}(c^{-2}).
 \label{eq:cor23}
\end{align}

Similarly, we obtain equations describing the light propagation in the counter-clockwise direction:
 {}
\begin{align}
c\Delta t_{42}&
\simeq  r_{42}+c^{-1}r_{42}\dot{ r}_{42} +{\cal O}(c^{-2}),
\label{eq:omcff4}\\
c\Delta t_{21}&
\simeq r_{21}+c^{-1}\Big(r_{42}\dot{ r}_{21}+(\vec r_{21}\cdot\dot{\vec r}_{41})-2 (\vec \omega \cdot \vec A_{412})\Big)
+{\cal O}(c^{-2}),
\label{eq:omc41ff}\\
c\Delta t_{14}&
\simeq r_{14}+ c^{-1}\dot{ r}_{14}\Big(r_{42}+r_{21}
\Big)+{\cal O}(c^{-2}).
 \label{eq:corc2ff2}
\end{align}

We may now combine the light transit times in the opposite directions forming a generalized Sagnac observable:
{}
 \begin{align}
\Delta L_{412}&=\ell^+-\ell^-=
c\big(t_{41}+t_{12}+t_{24}\big)-
c\big(t_{14}+t_{21}+t_{42}\big)=\nonumber\\
&\hskip -10pt
= \frac{1}{c}\Big\{
r_{12}(\dot{ r}_{42}-\dot{ r}_{41})-\dot{ r}_{12}(r_{42}-r_{41})+
r_{41}\dot{ r}_{42}-r_{42}\dot{ r}_{41}+(\vec r_{42}\cdot\dot{\vec r}_{41})-(\vec r_{41}\cdot\dot{\vec r}_{42})+4(\vec \omega \cdot \vec A_{412})\Big\}+{\cal O}(c^{-2}),
 \label{eq:genSag3}
  \end{align}
 where all the quantities on the right hand-side are taken at time $t_0$.

Similarly, one obtains expressions for the generalized Sagnac observables that describe light propagation along co- and contra-propagating paths along the other triangles of the evolving tetrahedron that share the same vertex \#4:
{}
 \begin{align}
\Delta L_{4ij}&= \Delta L^{\tt elast}_{4ij}+
\frac{4}{c}(\vec \omega \cdot \vec A_{4ij})+{\cal O}(c^{-2}), 
\label{eq:genSag3+}\\
\Delta  L^{\tt elast}_{4ij}&=\frac{1}{c}\Big\{
r_{ij}(\dot{ r}_{4j}-\dot{ r}_{4i})-\dot{ r}_{ij}(r_{4j}-r_{4i})+
r_{4i}\dot{ r}_{4j}-r_{4j}\dot{ r}_{4i}+(\vec r_{4j}\cdot\dot{\vec r}_{4i})-(\vec r_{4i}\cdot\dot{\vec r}_{4j})\Big\}+{\cal O}(c^{-2}).
 \label{eq:genSag3ij}
  \end{align}

Given the anticipated values of the inter-spacecraft velocities and accelerations within the constellation evaluated  and found to be $v_{ij}\simeq 0.20~{\rm m/s}$ and $a_{ij}\simeq 3.96\times 10^{-8}~{\rm m/s}^2$, correspondingly (see (\ref{eq:vel-acc}) and  Sec~\ref{sec:tetra-form} for details), we may evaluate the magnitude of $\Delta  L^{\tt elast}_{4ij}$ in (\ref{eq:genSag3ij}). Using these anticipated values, we estimate that  these terms will be on the order of $r_{ij}\dot r_{ij}/c\simeq 6.67\times 10^{-4}~{\rm m}$ and, given the anticipated sensitivity of the Sagnac interferometers in measuring the optical path difference (OPD), $\delta \Delta L_{4ij}=10$~pm, are large enough to be included in the model. 

We also developed the model that accounts for the relative accelerations between the vehicles and also velocity-dependent terms $\propto \Delta t^2$, thus improving the model (\ref{eq:r41})--(\ref{eq:cor}). Using this updated model, the relevant velocity- and acceleration-dependent terms were evaluated to be $r_{ij}\dot r_{ij}^2/c^2\simeq 4.45\times 10^{-13}~{\rm m}$ and $r_{ij}^2\ddot r_{ij}/c^2\simeq 4.41\times 10^{-13}~{\rm m}$, correspondingly.  Both of these contributions are small but, depending on the ultimate missions design, they may have to be included in the model.  Below, we limit our consideration only to the terms present in (\ref{eq:genSag3+})--(\ref{eq:genSag3ij}).

Note that all the terms in (\ref{eq:genSag3ij}) are available either from laser ranging measurements, $r_{ij}, \dot r_{ij}$ (and, thus, the area $\vec A_{4ij}$), or from Sagnac interferometry, $\Delta L_{4ij}$. Therefore, we may use these expressions to determine $\vec \omega$. For that, we consider the three triangles that share common vertex at spacecraft \#4, and, defining $\Delta \ell_{4ij}=\Delta L_{4ij}-\Delta L^{\tt elast}_{4ij}$, we have the following three equations to determine $\vec\omega$:
{}
\begin{align}
\Delta \ell_{4ij}= \frac{4}{c}\big(\vec\omega\cdot \vec A_{4ij}\big)\qquad {\rm with}\qquad
\vec A_{4ij}= {\textstyle\frac{1}{2}} [\vec r_{4i}\times \vec r_{4j}],
 \label{eq:sint412a}
\end{align}
which now accounts for the elasticity of the triangles and the oriented areas $\vec A_{4ij}$ are evaluated at the beginning of the Sagnac measurements. Next, combining, for instance,  three relevant equations one obtains
{}
\begin{align}
\Big(\vec\omega\cdot \big(\vec A^{\rm T}_{412}, \vec A^{\rm T}_{431}, \vec A^{\rm T}_{423}\big)\Big)={\textstyle\frac{1}{4}}c\, \Big(\Delta \ell_{412},\Delta \ell_{431},\Delta \ell_{423}\Big)
\qquad \Rightarrow \qquad
(\vec\omega\cdot {\vec A})={\textstyle\frac{1}{4}}c\, \Delta \vec \ell.
 \label{eq:sint412c+}
\end{align}

If ${\rm det} (\vec A)=\big(\vec A_{412}\cdot
[\vec A_{423}\times \vec A_{431}]\big)\not = 0$, the matrix $\vec A$ is invertible, yielding
 {}
\begin{align}
\vec A^{-1}&=\frac{1}{{\rm det} (\vec A)}\Big([\vec A_{423}\times \vec A_{431}]^{\rm T},[\vec A_{431}\times \vec A_{412}]^{\rm T},[\vec A_{412}\times \vec A_{423}]^{\rm T}\Big)\equiv
\Big(\frac{\tilde{\vec n}_{412}}{A_{412}},
\frac{\tilde{\vec n}_{423}}{A_{423}},
\frac{\tilde{\vec n}_{431}}{A_{431}}\Big),
 \label{eq:sint412c}
\end{align}
where $\tilde{\vec n}_{412}$, $\tilde{\vec n}_{423}$ and $\tilde{\vec n}_{431}$ are another set of the reciprocal base vectors composed from the  areal unit vectors $\vec n_{4ij}= [\vec r_{4i}\times \vec r_{4j}]/|[\vec r_{4i}\times \vec r_{4j}]|$ normal to the corresponding faces of the tetrahedron. Such a definition is similar to the one for vectors introduced in (\ref{eq:recipr}) with the resulting form being evident from the structure of (\ref{eq:sint412c}).

As a result, with the help of (\ref{eq:sint412c}), the angular velocity vector $\vec\omega$ may be determined from (\ref{eq:sint412c+}) as
{}
\begin{align}
\vec\omega={\textstyle\frac{1}{4}}c\, \big(\Delta \vec \ell \cdot {\vec A}^{-1}\big),
 \label{eq:sint412d}
\end{align}
providing the components of the angular velocity $\vec \omega$ that account for the elasticity of the tetrahedron:
{}
\begin{align}
\omega_x&={\textstyle\frac{1}{4}}c\frac{\big(\Delta \vec \ell \cdot \tilde {\vec n}_{412}\big)}{ A_{412}}, \qquad
\omega_y={\textstyle\frac{1}{4}}c\frac{\big(\Delta \vec \ell \cdot \tilde{\vec n}_{423}\big)}{A_{423}}, \qquad
\omega_z={\textstyle\frac{1}{4}}c\frac{\big(\Delta \vec \ell \cdot \tilde{\vec n}_{431}\big)}{A_{431}},
 \label{eq:sint41d+}
\end{align}
which is oriented with respect to the TCS defined by (\ref{eq:dg25da2=}).

Result (\ref{eq:sint41d+}) generalizes Sagnac observables in the case of an elastic tetrahedron. If precision laser ranging data are  available, it enables inertial navigation by providing accurate measurements of the angular velocity vector $\vec \omega$.  In other words, Sagnac interferometry conducted along the edges of  the three triangles with a common vertex within a tetrahedral configuration, combined with range measurements of the same edges enables determination of $\omega$.  That fact is important for the experiment as it provides a critical piece of information needed to determine  ${\rm tr} ({\bf T})_4$ from (\ref{eq:dgd+3}).

Expressions (\ref{eq:sint41d+}) allow us to consider the accuracy in determining the components of the angular velocity that may be provided by Sagnac observables. Generically, these expressions behave as
 {}
\begin{align}
\omega&\simeq c\frac{\Delta \ell }{2r_{4i}^2},
 \label{eq:sint41d+=}
\end{align}
which implies that, if the Sagnac interferometer will measure optical path differences with an accuracy of $\delta \Delta  \ell\simeq 10$~pm, this would enable a determination of the angular velocity of the constellation with respect to SSB frame with an accuracy of  $\delta \omega \lesssim c{\delta\Delta  \ell}/{2r_{4i}^2}\simeq 1.5\times 10^{-15}~{\rm s}^{-1}$, which satisfies the requirement set by (\ref{eq:dg2fg-om=}).  This demonstrates that the Sagnac observables are capable of a highly precise determination of the angular velocity, thus enabling the experiment.

\section{Tetrahedral formation on nearly identical eccentric orbits}
\label{sec:eccentric-orb} 

To describe the motion of the spacecraft with the tetrahedral configuration, we may rely either on a numerical analysis  or we may use an analytical approach considering that all the spacecraft follow nearly identical elliptic orbits. While in Sec.~\ref{sec:sim}, we discuss the results of a numerical analysis that provides detailed insights into the dynamics and possible real-world perturbations, below we develop analytical models needed to obtain the basic understanding of the relevant dynamics within the constellation. By using an analytical approach, one can get an insight into the fundamental dynamics of spacecraft in a tetrahedral configuration that will be helpful for mission design.

\subsection{Linear perturbations around a general reference orbit}
\label{sec:orb-pert}

Satellite-formation missions can be designed using two primary strategies: active control and natural formation \cite{Guzman-Schiff:2002,Clemente:2005}. In the active control method, satellites use thrusters to actively maintain or adjust their relative positions, ensuring constant or periodic geometrical configurations during the orbit. In contrast, the natural method designs the satellites' orbits such that they inherently achieve the desired formation based on scientific needs, without the continuous intervention of active controls. Here we adopt the natural formation approach relying on analytical methods to construct a tetrahedral formation \cite{Shestakov-etal:2019}.

Geometric techniques for the design of formation flying, relying on the analytical solution to Hill's equations, have been previously established and used in many efforts \cite{Carter-Humi:1987,Montenbruck-2005} and applied to define intended relative motions in orbits that are nearly circular. These methods establish understandable relationships between spacecraft, providing valuable understanding of relative motion. This facilitates the swift creation of satellite arrangements that fulfill specific mission criteria, such as achieving certain vehicle distances during perigee or apogee, ensuring minimal separation, or attaining a particular geometric pattern. Moreover, the outcomes derived from these geometric methods can effectively limit and guide numerical optimization approaches, leading to quicker attainment of optimal satellite configurations. 

In our case, our reference orbit has a significant eccentricity, rendering Hill's equations ineffective as they where developed for nearly circular reference orbits. Here, we consider the case of generic elliptic heliocentric orbits \cite{Inalhan:2002,Broucke:2003,Lane-Axelrad:2006} and use them to explore tetrahedral formation design and its temporal evolution.

To study the dynamical behavior of a tetrahedral configuration, we need to establish a set of geometrical relationships describing the relative motion of spacecraft in nearby eccentric orbits. For that, we follow \cite{Lane-Axelrad:2006} and assume that the primary vehicle, termed the reference center, follows an unperturbed, eccentric trajectory that is referred to as the reference orbit. (Note that quantities without subscripts refer to the reference center unless otherwise noted.) The reference orbit is completely described by the set of standard Keplerian orbital elements $\alpha =[a, e, i, \Omega, \omega, M_0]^{\rm T}$. Any of the vehicle, with the constellation is in a similar orbit with only a small change in orbital elements: $\alpha_i =\alpha+\Delta\alpha_i, i\in \{1,2,3\}$. The derivation requires the assumption that the orbital elements of all spacecraft within the constellation are similar, so that $\Delta \alpha_i\ll \alpha$; no assumptions about the eccentricity of the reference orbit are made, except for $e<1$.

Assuming that the spacecraft follow nearly identical elliptic orbits, their motion is governed by Kepler's laws.  However, since these are nearly identical elliptic orbits, the primary variations in the relative positions of the spacecraft will arise due to their phase differences in the orbits and not due to differences in the size or shape of the orbits.
For a purely analytical approach, one might consider only the central force and study the motion of the spacecraft with respect to the reference frame. The perturbations can be added later for a more accurate but complex model.

We consider the sensitivity of the reference orbit to small changes in $\alpha$ (as developed in \cite{Lane-Axelrad:2006}). For that, we consider the motion of spacecraft in the solar gravitational field neglecting the presence of other planets.  So, essentially the solar system barycentric frame (SSB) collapses to heliocentric inertial  (HCI) frame. In that HCI frame the position and velocity vectors of the reference center are related to $\alpha$ through the following expressions (see \cite{Montenbruck-2005}):   
{}
\begin{eqnarray}
\vec r&=&r
 \Bigg[
 \begin{aligned}[c]
\cos\Omega\cos\theta&-\sin\Omega\cos i \sin\theta \\
\sin\Omega \cos\theta&+\cos\Omega\cos i\sin\theta\\
&\sin i \sin\theta
  \end{aligned}\Bigg], \quad~
  \label{eq:r-vec}
\dot{\vec r}=
v
 \Bigg[
 \begin{aligned}[c]
-\cos\Omega\big(\sin\theta+e\sin\omega\big)&-\sin\Omega\cos i\big(\cos\theta+e\cos\omega\big) \\
-\sin\Omega \big(\sin\theta+e\sin\omega\big)&+\cos\Omega\cos i(\cos\theta+e\cos\omega\big)\\
\sin i \big(\cos\theta&+e\cos\omega\big)
  \end{aligned}\Bigg],~~~
  \label{eq:v-vec}
\end{eqnarray}
where
\begin{eqnarray}
r&=&\frac{a(1-e^2)}{1+e\cos\nu}, \qquad v=\sqrt{\frac{\mu}{a(1-e^2)}}.
  \label{eq:r-scal}
\end{eqnarray}

A natural coordinate frame for describing the relative motion of a spacecraft with respect to a reference center is the radial, in-track, cross-track (RIC) frame shown in Fig.~\ref{fig:RIC}, see details in \cite{Lane-Axelrad:2006}.  This is a non-inertial frame that moves with the reference center. Assuming that position $\vec r$ and velocity $\dot{\vec r}$ of the reference center are known, evaluated at the center, the fundamental directions of this frame are
{}
\begin{eqnarray}
\vec e_{\tt R}&=&\frac{\vec r}{r}, \quad~~ 
\frac{d\vec e_{\tt R}}{dt}=\frac{\dot{\vec r}-(\vec e_{\tt R}\cdot\dot{\vec r})\vec e_{\tt R}}{r}, \quad~~ 
 \vec e_{\tt C}=\frac{\big[\vec r\times \dot{\vec r}\big]}{\big |\big[\vec r\times \dot{\vec r}\big]\big |}, 
 \quad~~ 
 \frac{d\vec e_{\tt C}}{dt}=0, \quad~~ 
 \vec e_{\tt I}=\big[\vec e_{\tt C}\times \vec e_{\tt R}\big], \quad~~
\frac{d\vec e_{\tt I}}{dt}=\big[\vec e_{\tt C}\times \frac{d\vec e_{\tt R}}{dt}\big],~~~
  \label{eq:axis}
\end{eqnarray}
with the unit vector $\vec e_{\tt R}$ pointing radially outward from Sun's center and $\vec e_{\tt I}$ in the in-track direction along increasing true anomaly. This right-handed orthogonal reference frame is completed with $\vec e_{\tt C}$, pointing in the cross-track direction.

The reference orbit $\vec r$  is represented by the standard orbital elements $(a, e, i, \Omega, \omega, \theta)$, which correspond to the semi-major axis, eccentricity, inclination, right ascension of the ascending node, argument of periapsis, and true anomaly, \cite{Montenbruck-2005}. Also, the radius $r=a(1-e^2)/(1+e\cos\nu)$ and the angular velocity of the formation center given as usual by $\dot{\theta} = {n(1+e\cos\theta)^2}/{(1-e^2)^\frac{3}{2}}$, with $\theta=\nu+\omega$. As defined, the RIC frame is non-inertial, undergoing rotation with the angular frequency vector $\vec \omega_{\tt RIC}$ directed along the angular momentum vector.

The sensitivity matrix $\vec S_{\rm HCI}$ is constructed by assembling the partials of each component of $\vec r$ with respect to each orbital element in $\alpha$ with details  given in Appendix~\ref{sec:partials}. As a result, at the reference center, the transformation matrix $\vec R$ relating the HCI frame to the RIC frame is given by
{}
\begin{eqnarray}
\vec R&=&
 \Bigg[
 \begin{aligned}[c]
\cos\Omega\cos\theta-\sin\Omega\cos i \sin\theta, ~~~~
&\sin\Omega \cos\theta+\cos\Omega\cos i\sin\theta, &\sin i \cos\theta\\
-\cos\Omega\sin\theta-\sin\Omega\cos i\cos\theta,~~-&\sin\Omega\sin\theta+\cos\Omega\cos i \cos\theta, &\sin i \cos\theta\\
\sin\Omega\sin i, ~~-&\cos\Omega\sin i,  &\cos i~~~~~~
  \end{aligned}\Bigg].
  \label{eq:mat-T}
\end{eqnarray}

The partials of $\vec r$ with respect to the set of orbital parameters $\alpha$ can be expressed in RIC by premultiplying $\vec S_{\rm HCI}$ (Appendix~\ref{sec:partials}) by $\vec  R$. This results in the RIC sensitivity matrix $\vec S_{\rm RIC}$,
{}
\begin{equation}
\vec S_{\rm RIC}=\vec R \, \vec S_{\rm HCI} =
\begin{bmatrix}
\frac{r}{a}-\frac{3n(t-t_0)e\sin\nu}{2\sqrt{1-e^2}} ~~~
&-a\cos\nu& 0&0&0&\frac{ae\sin\nu}{\sqrt{1-e^2}}\\[10pt]
-\frac{3an(t-t_0)\sqrt{1-e^2}}{2r}
&\Big(a+\frac{r}{1-e^2}\Big)\sin\nu &0 &r\cos i &r& \frac{a^2}{r}\sqrt{1-e^2}\\
0&0&r\sin\theta &-r\sin i\cos\theta & 0& 0
\end{bmatrix}.
  \label{eq:mat-Sric}
\end{equation}
Thus, $\vec S_{\rm RIC}$ is a mapping relating orbital element differences between the reference center and a spacecraft within the tight formation (the column space of $\vec S$) to radial, in-track, and cross-track position differences (the row space of $\vec S$).

Given the HCI position and velocity vectors of the reference center and  spacecraft with the constellation, one can compute the curvilinear representation of the relative position of the the $i$-th spacecraft with respect to the reference center (the origin of the RIC). Evaluated at the center, the sensitivity of the curvilinear coordinates $x_i$, $y_i$, and $z_i$ to orbit element differences is equal to ${\vec S}_{\rm RIC}$ given in Eq.~(\ref{eq:mat-Sric}). Thus, for small orbital element differences, the resulting $x_i, y_i$, and $z_i$ coordinates of the $i$-spacecraft are computed by $\vec S_{\rm RIC}\Delta\alpha_i$. This yields
{}
\begin{eqnarray}
x_i&=&\Big[
\frac{r}{a}-\frac{3n(t-t_0)e\sin\nu}{2\sqrt{1-e^2}}\Big]\Delta a_i-a\cos\nu\Delta e_i +\frac{ae\sin\nu}{\sqrt{1-e^2}}\Delta M_i,\nonumber\\
y_i&=&-\frac{3an(t-t_0)\sqrt{1-e^2}}{2r}\Delta a_i+\Big(a+\frac{r}{1-e^2}\Big)\sin\nu\Delta e_i +r\Big(\cos i\Delta \Omega_i+\Delta \omega_i\Big)+ \frac{a^2}{r}\sqrt{1-e^2}\Delta M_i,\nonumber\\
z_i&=&
r\sin\theta\Delta i_i -r\sin i\cos\theta \Delta \Omega_i,
  \label{eq:solut-pos-A}
\end{eqnarray}
where $n=\sqrt{\mu_\odot/a^3}$ is the natural frequency of the reference orbit.

The velocity equations are obtained by taking the time derivatives of (\ref{eq:solut-pos-A}):
{}
\begin{eqnarray}
\dot x_i=\frac{dx_i}{dt}&=&-\Big[
\frac{ne\sin\nu}{2\sqrt{1-e^2}}+\frac{3a^2}{r^2}n^2(t-t_0)e\cos\nu \Big]\Delta a_i+n\sin\nu\sqrt{1-e^2}\Big(\frac{a^3}{r^2}\Big)\Delta e_i +en\cos\nu \Big(\frac{a^3}{r^2}\Big)\Delta M_i,\nonumber\\
\dot y_i=\frac{dy_i}{dt}&=&\Big[\frac{3a^2}{2r^2}n^2(t-t_0)e\sin\nu-\frac{3a}{2r}n\sqrt{1-e^2}\Big]\Delta a_i+\Big[n\sqrt{1-e^2}\Big(1+\frac{r}{p}\Big)\Big(\frac{a^3}{r^2}\Big)\cos\nu+\frac{aen\sin^2\nu}{(1-e^2)^\frac{3}{2}}\Big]\Delta e_i +\nonumber\\
&&+\frac{aen\cos i\sin\nu}{\sqrt{1-e^2}}\Delta\Omega_i+\frac{aen\sin\nu}{\sqrt{1-e^2}}\Delta \omega_i- en\sin\nu\Big(\frac{a^3}{r^2}\Big)\Delta M_i,\nonumber\\
\dot z_i=\frac{dz_i}{dt}&=&\frac{an}{\sqrt{1-e^2}}\Big(\big(\cos\theta+e\cos\omega\big)\Delta i_i
+\sin i\big(\sin\theta+e\sin\omega\big)\Delta \Omega_i\Big).
  \label{eq:soluvel-A}
\end{eqnarray}
We use these equations to study the formation and evolution of a tetrahedral spacecraft configuration.

The distinction between time-independent analytical expressions and time-dependent numerical solutions provides a multifaceted understanding of the system's dynamics. Analytical models, like (\ref{eq:solut-pos-A})--(\ref{eq:soluvel-A}), offer insights into fundamental behavior by emphasizing key parameters such as the true anomaly $\nu$. Meanwhile, the numerical solutions shown in Sec. \ref{sec:sim}, will present a detailed temporal evolution, capturing intricate dynamics and possible real-world perturbations. Both approaches are invaluable for a holistic understanding of the spacecraft's tetrahedral configuration. 

\subsection{Setting up a representative spacecraft formation}
\label{sec:tetra-form}

Positioning satellites in elliptic orbits around the Sun with high eccentricity provides a varying distance $r$, which is crucial for detecting our signals of interest that are expected to exhibit distance dependence.  Circular orbits would sample nearly constant background, potentially missing the signals. Thus, to effectively measure these new effects, a tetrahedral configuration of four smallsats on elliptic orbits is considered. Furthermore, each satellite in this configuration will have nearly the same semi-major axis but will require precise initial positioning to ensure their relative phasing and the effective maintenance of a tetrahedral formation in terms of propulsion, power and communication. Over time, natural perturbations can disrupt this arrangement, necessitating onboard propulsion systems or other correction mechanisms to uphold the desired geometric constraints.

Consider a reference frame with its origin at a particular point within the tetrahedral formation. In this frame, if we know the position of one spacecraft, and we know the phase differences between the spacecraft, using (\ref{eq:solut-pos-A})--(\ref{eq:soluvel-A}), we can determine the positions and velocities of the other spacecraft in the formation. We use that approach below to set up a representative formation that allows us to learn on a dynamical behavior within the tetrahedron. 

Based on (\ref{eq:solut-pos-A}), geometrical relationships that describe the relative motion in eccentric orbits are established. Stable formations with no drift are of primary interest (as we are interested to explore the existence of passive orbits with no active control), and thus, the secular growth in the separation between the spacecraft is eliminated by constraining the energy of the orbits to be equal, that is, $\Delta a_i =0$. This leaves
{}
\begin{eqnarray}
x_i&=&-a\cos\nu\Delta e_i +\frac{ae\sin\nu}{\sqrt{1-e^2}}\Delta M_i,\nonumber\\[-6pt]
y_i&=&\Big(a+\frac{r}{1-e^2}\Big)\sin\nu\Delta e_i +r\Big(\cos i\Delta \Omega_i+\Delta \omega_i\Big)+ \frac{a^2}{r}\sqrt{1-e^2}\Delta M_i,\nonumber\\
z_i&=&
r\sin\theta\Delta i_i -r\sin i\cos\theta \Delta \Omega_i.
  \label{eq:solut-pos-intrack}
\end{eqnarray}
Other approaches for constraining the secular growth between the vehicles are needed when perturbations cause additional secular drifts. However, in unperturbed orbits, setting $\Delta a_i =0$ is sufficient.

Eq.~(\ref{eq:solut-pos-intrack}) can be used to study various spacecraft formation designs and their temporal evolutions. We have studied several   special cases of spacecraft formations, including: i)  in-track, ii) radial/in-track, and iii) radial/in-track/cross-track formations.  Many other formations exist and may be studied with expressions (\ref{eq:solut-pos-intrack}), however, these three were sufficient to establish our understanding needed to investigate tetrahedral formations of interest. Below, we use the third type of these formations that offers the most general approach to set up a tetrahedral formation.

In that regard, we study no-drift formations with spacecraft moving in all there dimensions in the RIC frame. For that, we allow for small cross-track motion, thus requiring that only $\Delta M_i =\Delta\omega_i=0$. In this case, (\ref{eq:solut-pos-intrack}) results in
{}
\begin{eqnarray}
x_i=-a\cos\nu\Delta e_i, \quad
y_i=\Big(a+\frac{r}{1-e^2}\Big)\sin\nu\Delta e_i +r\cos i\Delta \Omega_i, \quad
z_i=
r\sin\theta\Delta i_i - r\sin i\cos\theta\Delta \Omega_i.~~
  \label{eq:solut-pos-CFT}
\end{eqnarray}
This leaves us three design parameters to specify for a cross-track formation: $\Delta e_i, \Delta\Omega_i$, and $\Delta i_i$.

To determine the constants involved in (\ref{eq:solut-pos-CFT}), we can set the formation at the perigee by selecting $\nu=0$. Remembering that $\theta=\nu+\omega$, Eqs.~(\ref{eq:solut-pos-CFT}) are matched to their initial values yielding
{}
\begin{eqnarray}
\Delta e_i= -\frac{x_{i0}}{a},
\qquad
\Delta \Omega_i= \frac{y_{i0}}{a(1 -e)\cos i},
\qquad \Delta i_i= \frac{z_{i0}+y_{i0} \tan i\cos\omega}{a(1 -e)\sin \omega},
  \label{eq:solut-con3d}
\end{eqnarray}
where $\vec r_{i0}=(x_{i0},y_{i0},z_{i0})$ are the initial positions of the $i$-th spacecraft. We can safely assume that the reference orbit is within the ecliptic plane (i.e., $i=0$) and the  perigee is located at $\omega=\pi/2$.  In this case,  (\ref{eq:solut-con3d}) yields the following 
{}
\begin{eqnarray}
\Delta e_i= -\frac{x_{i0}}{a},
\qquad
\Delta \Omega_i= \frac{y_{i0}}{a(1 -e)},
\qquad \Delta i_i = \frac{z_{i0}}{a(1 -e)}.
  \label{eq:solut-con3}
\end{eqnarray}

As a result, the position of the $i$-th vehicle identifies a relevant  formation in the RIC frame given as follows
{}
\begin{eqnarray}
x_i=x_{i0}\cos\nu, \qquad
y_i=-\Big(1+\frac{r}{a(1-e^2)}\Big) x_{i0} \sin\nu +\frac{r}{a(1-e)} y_{i0}, \qquad
z_i=
\frac{r}{a(1-e)}z_{i0}\cos\nu.
  \label{eq:solut-pos-CFT-pos1}
\end{eqnarray}

We also need to evaluate the relative velocities within the constellation. For that, using (\ref{eq:soluvel-A}) for the same conditions that where used to derive solution (\ref{eq:solut-pos-CFT-pos1}), namely $\Delta a_i=\Delta M_i =\Delta\omega_i=0$, as well as $\omega=\pi/2$, $i=0$, and using initial conditions (\ref{eq:solut-con3}), we have the following velocity components:
{}
\begin{eqnarray}
\dot x_i &=&-n\sin\nu\sqrt{1-e^2}\Big(\frac{a^2}{r^2}\Big)x_{i0},
\qquad\qquad
\dot z_i
=-\frac{n \sin\nu}{\sqrt{1-e^2}(1 -e)}z_{i0},
\nonumber\\
\dot y_i
&=&-\Big[n\sqrt{1-e^2}\Big(1+\frac{r}{p}\Big)\Big(\frac{a^2}{r^2}\Big)\cos\nu+\frac{en\sin^2\nu}{(1-e^2)^\frac{3}{2}}\Big]x_{i0} +
\frac{en\sin\nu}{\sqrt{1-e^2}(1 -e)}y_{i0}.
  \label{eq:soluvel2}
\end{eqnarray}

Results (\ref{eq:solut-pos-CFT-pos1})--(\ref{eq:soluvel2}) suggest that  not only the separation between the vehicles changes as they move on their elliptic and nearly identical orbits, their mutual orientation also periodically varies. In addition, the entire tetrahedron rotates with the natural frequency of the orbit.  

We may use the results above to evaluate the dynamical behavior within the constellation. To do that, we consider the reference orbit with the semi-major axis of $a=1$~AU, eccentricity $e=0.6$ (see Table~\ref{tb:sim-params}). For such a configuration, the natural orbital frequency is $n=\sqrt{\mu_\odot/a^3}\simeq 1.99\times 10^{-7}~{\rm s}^{-1}$. Considering the inter-spacecraft separation of $r_{ij}\sim 10^3$~km, the  nominal relative velocities, $v_{ij}$, and accelerations, $a_{ij}$, within the constellation are estimated to be
{}
\begin{eqnarray}
v_{ij}\sim n r_{ij}\simeq 0.20~{\rm m/s},
\qquad 
a_{ij}\sim n^2 r_{ij}\simeq 3.96\times 10^{-8}~{\rm m/s}^2.
  \label{eq:vel-acc}
\end{eqnarray}

We note that, as evidenced by the form of  (\ref{eq:solut-pos-CFT-pos1}) and (\ref{eq:soluvel2}), even in this rather simple case, in addition to the natural frequency, $n$, there will be other frequencies present in $v_{ij}$ and $a_{ij}$.  As a result, as the spacecraft move in the their ecliptic orbits around the Sun, the values (\ref{eq:vel-acc}) will be modulated and amplified, resulting in changes that are small but important to be accounted for when considering a realistic mission architecture. Therefore, below, we use results (\ref{eq:vel-acc}) as nominal representative values when addressing the  error budget and mission design. 

\subsection{Tetrahedral formation on an eccentric orbit}
\label{sec:tetra-form-e}

A natural basis for inertial measurements and scientific observations is the orbiting (non-inertial) reference frame, fixed to the formation center. To describe the relative motion of the spacecraft within the constellation, one needs to choose a convenient reference point. There are two options  for such a choice, namely
{}
\begin{enumerate}[i).]

\item Mesocenter of the tetrahedron \cite{Guzman-Schiff:2002}: Choosing mesocenter  (i.e., mean position or centroid) may seem convenient from the dynamical standpoint; however, the position of this point is not directly measured, observed and/or otherwise maintained. In fact, such a position would have to be computed using spacecraft orbital positions, their masses, and rotation states. As most of the relevant trajectory information will be provided by the DSN, it may not be of the highest accuracy needed for the experiment, making this option suboptimal.

\item Any spacecraft within the constellation: Choosing one of the spacecraft as the reference point has a practical advantage as individual spacecraft orbits are going to be available from DSN and, thus, are directly observed.  Furthermore, such a choice  would allow us to benefit from laser ranging that will provide us with highly precise inter-spacecraft range measurements. Also, such a choice is consistent with the relative nature of the measurements which are more convenient to describe relying on the set of measurements taken on-board.
\end{enumerate}

Based on the arguments above, one can choose one of the spacecraft to be the origin of the orbital coordinate system, for convenience, placing it at spacecraft \#4. Note that such a coordinate system may be introduced at any spacecraft within the constellation.

To describe the tetrahedron formation in the case when one of the vehicles is on the reference orbit, we use the solution (\ref{eq:solut-pos-CFT-pos1}). We choose spacecraft \#4 to be on the reference orbit and set up a regular tetrahedron. 

There are many ways to set up such a configuration. As an example, we consider a LISA-like tetrahedron formation (with one of vehicles being on the reference orbit), and the fourth vehicle completing a regular tetrahedron. This can be done by choosing the following initial positions of the vehicles (where, again, $\vec r_{04} = (0,0,0)$):
{}
\begin{eqnarray}
\vec r_{01} = \Big({\textstyle\frac{\sqrt{3}}{2}}\ell_0, \,{\textstyle\frac{1}{2}}\ell_0,\,0\Big),
 \qquad
\vec r_{02} = \Big({\textstyle\frac{\sqrt{3}}{2}}\ell_0, \,-{\textstyle\frac{1}{2}}\ell_0,\,0\Big),
\qquad
\vec r_{03} = \Big({\textstyle\frac{1}{\sqrt{3}}}\ell_0, 0,\,{\textstyle\sqrt{\frac{2}{3}}}\ell_0\Big).
\label{eq:pos0-LISA+}
\end{eqnarray}

As a result, with the help of  (\ref{eq:solut-pos-CFT-pos1}),  one configures a tetrahedral formation:
{}
\begin{eqnarray}
\vec r_{1,2} &=& {\textstyle\frac{\sqrt{3}}{2}}\ell_0\Big\{\cos\nu, ~
-\Big(1+\frac{r}{a(1-e^2)}\Big) \sin\nu \pm {\textstyle\frac{1}{\sqrt{3}}}\frac{r}{a(1-e)}, ~ 0\Big\},\nonumber \\
\vec r_3 &=& {\textstyle\frac{1}{\sqrt{3}}}\ell_0\Big\{\cos\nu, ~
-\Big(1+\frac{r}{a(1-e^2)}\Big) \sin\nu,
~\frac{\sqrt{2}r}{a(1-e)}\cos\nu\Big\},
\label{eq:ctr33-LISA+}
\end{eqnarray}
where $`+'$ sign is for spacecraft \#1 and $`-'$ sign for spacecraft \#2.

Clearly, many other tetrahedral configurations exist and must be studied and optimized for the ultimate mission. Opting for the LISA-like configuration, as defined by equations (\ref{eq:pos0-LISA+})--(\ref{eq:ctr33-LISA+}), serves a dual purpose. First, it allows the focus to be on a well-studied configuration with known desirable properties, which facilitates benchmarking and validates the analytical and numerical methods used. Second, since both the configurations considered were developed using the same foundational approach embodied in equation (\ref{eq:solut-pos-CFT-pos1}), they are likely to exhibit closely related dynamical behavior. This choice thus permits a detailed exploration of a subset of the parameter space without loss of generality, thereby offering valuable insights for the mission design while ensuring computational and analytical efficiency.

\begin{figure}[t!]
\begin{minipage}[b]{.46\linewidth}
\rotatebox{90}{\hskip 5pt  Normalized volume, $V(\nu)/V(0)$}
\includegraphics[width=0.95\linewidth]{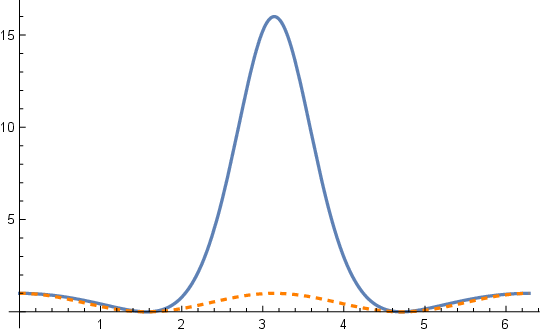}
\rotatebox{0}{\hskip 30pt  True anomaly $\nu\in[0,2\pi]$}
\end{minipage}
\,
\begin{minipage}[b]{.46\linewidth}
\rotatebox{90}{\hskip 5pt  Normalized volume, $V(\nu)/V(0)$}
\includegraphics[width=0.95\linewidth]{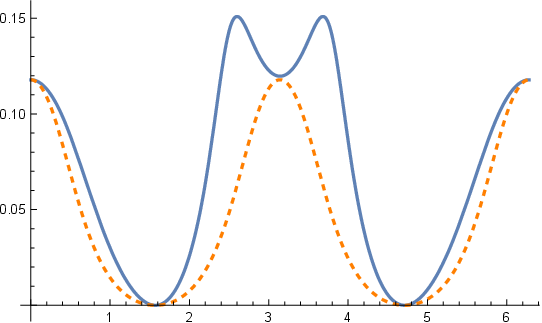}
\rotatebox{0}{\hskip 40pt  True anomaly $\nu\in[0,2\pi]$}
\end{minipage}
\vskip -2pt
\caption{\label{fig:vol} Left: Evolution of the tetrahedron volume as a function of true anomaly $\nu$ and eccentricity, $e$, as given by (\ref{eq:vol-det-LISA+}). Vertical axis shows the volume compared to its initial value at the perigee,  $V(\nu)/V(0)$, where formation was initiated. Right: Changes in the normalized volume $({\vec n}_{41}\cdot [\vec n_{42}\times \vec n_{43}])=V(\nu)/ r_{41} r_{42} r_{43}$.
Horizontal axis: $\nu\in[0,2\pi]$. Dashed:  $e=0$; solid:  $e=0.6$.
}
\vskip 12pt
\begin{minipage}[b]{.46\linewidth}
\rotatebox{90}{\hskip 7pt  Normalized range $r_{41}(\nu)/r_{41}(0)$}
\includegraphics[width=0.95\linewidth]{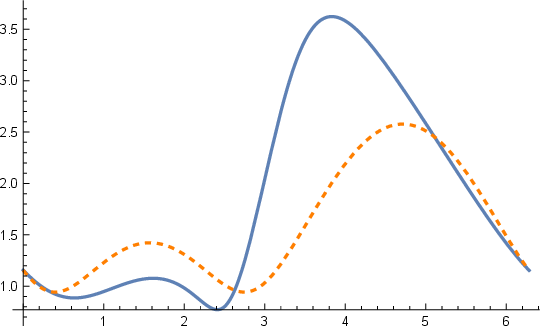}
\rotatebox{0}{\hskip 40pt  True anomaly $\nu\in[0,2\pi]$}
\end{minipage}
~\,
\begin{minipage}[b]{.46\linewidth}
\rotatebox{90}{\hskip 20pt  Pointing angle $\theta_{41}(\nu)$, deg}
\includegraphics[width=0.95\linewidth]{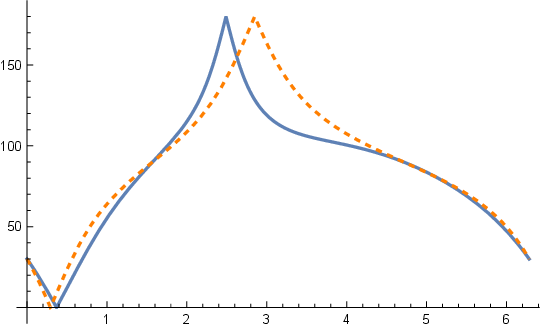}
\rotatebox{0}{\hskip 40pt  True anomaly $\nu\in[0,2\pi]$}
\end{minipage}
\vskip -2pt
\caption{\label{fig:r41} Left: Distance $r_{41}$ as a function of $\nu$ and eccentricity, $e$, as given by (\ref{eq:r41}). Vertical axis shows the distance compared to its initial value at perigee,  $r_{41}(\nu)/r_{41}(0)$. 
\label{fig:n41ang}
Right: Pointing angle  $\theta_{41}(\nu)$, as a function of $\nu$ and  eccentricity, $e$, as given by (\ref{eq:nn41}). Vertical axis: angle in degrees. Horizontal axis is $\nu\in[0,2\pi]$. For both plots: dashed:  $e=0$, solid:  $e=0.6$.
}
\end{figure}

The oriented volume of the tetrahedron formed by the four satellites is a fundamental descriptor of its spatial configuration. Using the defined relative vectors  $\vec r_{ij}=\vec r_{j}-\vec r_{i}$, the  volume can be expressed in terms of the scalar triple product. Using the specific values from (\ref{eq:ctr33-LISA+}), one can compute the magnitude and sign of the volume, which respectively give the volume's size and the orientation (chirality) of the tetrahedral configuration:
{}
\begin{eqnarray}
V &=&
{\textstyle\frac{1}{6}}(\vec r_{41}\cdot [\vec r_{42}\times \vec r_{43}])\equiv {\textstyle\frac{1}{6}}\det\Big[\vec r_{41}, \vec r_{42}, \vec r_{43}\Big]=
-{\textstyle\frac{1}{6\sqrt{2}}}  \ell_0^3\frac{(1+e)^2 \cos^2\nu}{(1+e\cos\nu)^2}.
  \label{eq:vol-det-LISA+}
\end{eqnarray}
Fig.~\ref{fig:vol}  shows the characteristic behavior of the volume (\ref{eq:vol-det-LISA+}) that takes four principal values, namely at perigee, $\nu=0$, it has its nominal value; at  $\nu={\textstyle\frac{\pi}{2}},{\textstyle\frac{3\pi}{2}}$ it collapses to zero; and at $\nu=\pi$ it reaches its maximum value, all given below
\begin{equation}
V(0) =
{\textstyle\frac{1}{6\sqrt{2}}}  \ell_0^3, \qquad
V({\textstyle\frac{\pi}{2}},{\textstyle\frac{3\pi}{2}}) =
0,  \qquad
V(\pi)={\textstyle\frac{1}{6\sqrt{2}}}  \ell_0^3\frac{(1+e)^2}{(1-e)^2}.
\label{eq:vol1}
\end{equation}
Note that for $e=0.6$, as the vehicles move on their orbits between perigee and apogee, the volume increases by a factor of $(1+e)^2/(1-e)^2=16$, as shown in Fig.~\ref{fig:vol} (left). (Note that in the case when $e =0$, the entire tetrahedron preserves its shape and initial volume while rotating with the orbital frequency.)

In addition, we need to understand the behavior of the normalized unit volume that enters the denominators in the reciprocal coordinate basis (\ref{eq:recipr}). The quantity we are interested in is $v=({\vec n}_{41}\cdot [\vec n_{42}\times \vec n_{43}])$, which may be computed from (\ref{eq:vol-det-LISA+}) as   $v(\nu)=V(\nu)/ r_{41} r_{42} r_{43}$. Because of the significant variability in the lengths of the tetrahedral edges, this quantity exhibits significant changes as the constellation moves in its heliocentric orbit, see Fig.~\ref{fig:vol} (right). Understanding the behavior of this quantity is important for the mission designs as we want to know the orbital regions where $v$  vanishes as the overall solution (\ref{eq:dgd+3}) experiences large variability, as will be shown in Fig.~\ref{fig:alt1}. 
  
Expressions (\ref{eq:ctr33-LISA+}) allow us to evaluate the behavior of the relative vectors between the spacecraft. With $\vec r_4=0$, we have  $\vec r_{4i}=\vec r_{i}$ while the remaining vector differences are readily computed as $\vec r_{ij}=\vec r_{j}-\vec r_{i}$ allowing us to study the internal dynamics of the tetrahedral configuration via the displacement between individual spacecraft pairs. To appreciate the dynamics of the entire tetrahedral structure we may examine either of the vectors to infer the overall behavior and stability of the tetrahedral formation. Taking, for instance, $\vec r_{41}$, we model the range between the two vehicles as
{}
\begin{eqnarray}
 r_{41} &=& {\textstyle\frac{\sqrt{3}}{2}}\ell_0\Big[\cos^2\nu+
\Big({\textstyle\frac{1}{\sqrt{3}}}\frac{r}{a(1-e)}-\Big(1+\frac{r}{a(1-e^2)}\Big) \sin\nu\Big)^2\Big]^\frac{1}{2}.
\label{eq:r41a}
\end{eqnarray}
Fig.~\ref{fig:r41} (left) shows the range evolution as the spacecraft move in their orbit, indicating that for $e=0.6$, it periodically increases by $\sim 3.83$ times. In Sec.~\ref{sec:sim}, we further explore this evolution with numerical simulations.

Similarly, we can also analyze the behavior of the pointing angle between the vehicles 4 and 1. For that, we model pointing with the unit vector $\vec n_{41}$ as usual:
{}
\begin{eqnarray}
\vec n_{41} &=& \frac{\vec r_{41}}{ r_{41}}=
\Big(\cos \theta_{41}, ~\sin \theta_{41}, ~0\Big).
\label{eq:nn41}
\end{eqnarray}
The knowledge of the pointing vector is important as it informs the technical aspects of designing the laser interferometric ranging systems for GDEM. In particular, this quantity determines the ranges of the angular articulation for the three small optical telescopes that will be positioned at each of the vehicle to enable $r_{ij}$ measurements. In addition, the same system will be used to provide attitude $\vec \omega$ via Sagnac measurements. 

Fig.~\ref{fig:n41ang} (right), shows the evolution of the pointing angles with respect to the true anomaly. It emphasizes the dynamical behavior of the tetrahedral configuration, particularly highlighting the significant variation in the pointing angle between vehicles 4 and 1. While this variation is noteworthy, its gradual nature provides an opportunity for mitigation. The slow rate of change implies that with astute instrumental calibrations and effective mission design, it's possible to address and compensate for these angle excursions, ensuring the robustness of the spacecraft formation and the integrity of the mission's objectives. In Sec.~\ref{sec:sim}, we further explore this evolution with numerical simulations.  

Figs.~\ref{fig:vol}--\ref{fig:n41ang} show the dynamic nature of the tetrahedral constellation. It is evident that the constellation doesn't maintain a static formation. Rather, it behaves much like an elastic body—stretching, compressing and twisting, and concurrently experiencing kinematic rotations. Such behavior is intrinsically tied to the satellites' orbital dynamics and the gravitational interactions that govern their motion. Given this dynamic nature, it is vital to consider these characteristics in mission planning. For the mission's success, accounting for these elastic behaviors and rotations is essential, these influencing both the calibration of instruments and the overall mission design.

Linear approximation techniques have been employed here to determine essential observables and establish constraints for achieving a tetrahedral configuration. Initial findings indicate that measurements based on this approach yield an error margin of just 0.1\% (due to the terms in the model beyond those present in (\ref{eq:eq-ff})), offering a notable reduction in uncertainties when contrasted with traditional methods. Our next step involves a more in-depth exploration of this dynamics through numerical simulations.

\subsection{Numerical Simulations}
\label{sec:sim}

\begin{table*}[t!]
\vskip-15pt
\caption{Selected mission parameters of the GDEM mission used in the simulations.
\label{tb:sim-params}}
\begin{tabular}{|l|c|c|}\hline
Parameter  &Symbol  &Value\\\hline\hline
Semimajor axis &$a$ &\phantom{0}1~AU\phantom{/s}\\
Orbital eccentricity&$e$ &\phantom{00}$\sim$0.6\phantom{/s}\\
Heliocentric velocity      & $v_{ j} \simeq \sqrt{{\mu_\odot/a}}$  &\phantom{.} 29.78~km/s\phantom{}\\
Mean orbital frequency & $n \simeq \sqrt{\mu_\odot/a^3}$  &\phantom{0}$1.99\times 10^{-7}~{\rm s}^{-1}$\phantom{k}\\
Heliocentric acceleration & $a_{0} =\mu_\odot/a^2$~\, & \phantom{0.} $5.93\times 10^{-4}$~{\rm m/s}$^2$\phantom{k}\\
Inter-spacecraft range &$r_{ij}$ &\phantom{0}$10^3$~km\phantom{/s}\\
Inter-spacecraft range rate &$\dot r_{ij}\simeq n r_{ij}$ &\phantom{0}0.20~{\rm m/s}\phantom{k}\\
Relative spacecraft acceleration &$\ddot r_{ij}\simeq n^2 r_{ij}$&\phantom{00}$3.96\times 10^{-8}$~{\rm m/s}$^2$
\phantom{k}\\\hline
\end{tabular}
\caption{Top level instrumental requirements for the GDEM mission, along with corresponding symbols used in the text.
\label{tb:inst-params}}
\begin{tabular}{|l|c|c|}\hline
Parameter  &Symbol  &Value\\\hline\hline
Laser ranging &$\delta {r}_{4i}$ &\phantom{0}$1\times 10^{-11}~{\rm m}$\phantom{/s}\\
Range-rates &$\delta \dot{r}_{4i}$ &\phantom{0}$1\times 10^{-5}~{\rm m/s}$\phantom{/s}\\
Line-of-sight accelerations &$\delta \ddot{r}_{4i}$ &\phantom{00}$1\times 10^{-15}~{\rm m/s}^{2}$\phantom{/s}\\
AI, as inertial sensor &$\delta f_{4i}$ &\phantom{0}$1\times 10^{-15}~{\rm m/s}^{2}$\phantom{/s}\\
Sagnac observable  &$\delta \omega$&\phantom{0.}$1.5\times 10^{-15}~{\rm s}^{-1}$\phantom{k}\\\hline
\end{tabular}
\end{table*}
\vskip-0pt

To validate the analysis and results that were discussed in the previous sections, we developed a simulation software. Implemented in HTML and JavaScript (to be expanded and improved in further studies), the software offers a simple visualization of the evolving tetrahedron configuration while at the same time calculating the trace of the GGT, as well as the same trace, recovered from intersatellite range and generalized Sagnac observables.

\begin{figure}
\begin{minipage}[b]{.96\linewidth}
\rotatebox{90}{\hskip 73pt  ${\rm tr}(\vec T)$}
\hskip -5pt 
\includegraphics[]{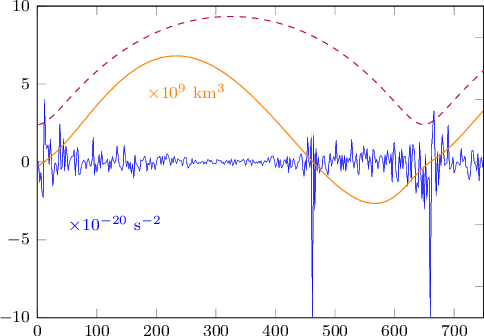}~\includegraphics[]{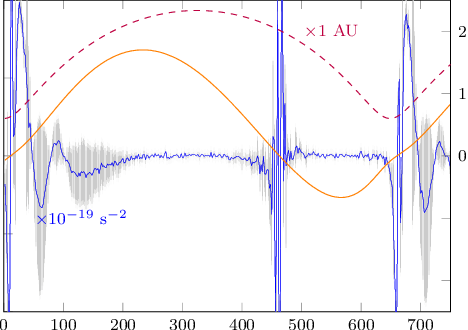}
\rotatebox{90}{\hskip 30pt  Heliocentric distance, AU}
\rotatebox{0}{\hskip 20pt  Days \hskip 210pt Days}
\end{minipage}
\caption{\label{fig:alt1} Simulation of the trace of the GGT.
Left: results for a specific reference orbit with a perihelion of 0.6~AU and eccentricity $e\sim 0.5909$, with the satellites initially forming a tetrahedron with $\sim 1,000$~km edges. 
Right: shows the values for the trace of the GGT that are derived from the actual ranges between satellites and the generalized Sagnac observables (see Table~\ref{tb:inst-params}).
The constellation volume is also shown (orange solid line). Dashed purple line shows the heliocentric distance, referencing the secondary vertical axis. Horizontal axis: days since perihelion passage.}
\vskip 6pt
\begin{minipage}[b]{.96\linewidth}
\rotatebox{90}{\hskip 73pt  ${\rm tr}(\vec T)$}
\hskip -5pt 
\includegraphics[]{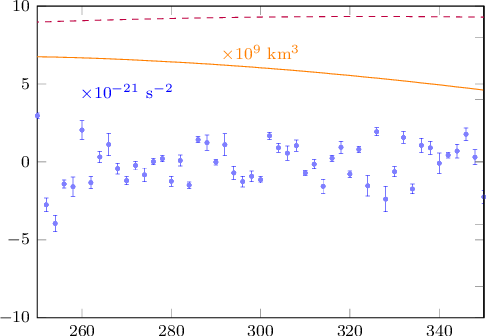}~\includegraphics[]{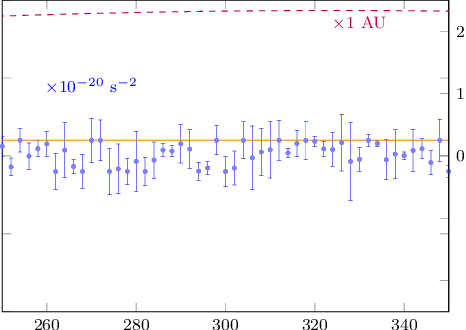}
\rotatebox{90}{\hskip 33pt  Heliocentric distance, AU}
\rotatebox{0}{\hskip 22pt  Days \hskip 200pt Days}
\end{minipage}
\caption{\label{fig:alt2} Details from the simulation shown in Fig.~\ref{fig:alt1} from a region of maximum sensitivity.
}
\end{figure}

In our simulation, the gravitational influence is limited exclusively to the Sun's gravity field. While gravitational effects from other bodies in the solar system do exist, their impact on our results is negligible\footnote{Note that GDEM relies on differential observables taken at the nominal inter-spacecraft separation of $r_{ij}=1000$\,km, see Table~\ref{tb:inst-params}.} This is due to the vacuum Poisson equation, i.e., (\ref{eq:ENeq}), with $\rho=0$. Consequently, while the gravitational field influences the satellites' orbits, it does not directly affect the term ${\rm tr}(\vec T)$. This effect is only through the approximations previously discussed. The contribution from the known solar system bodies other than the Sun to these terms is minimal and can be considered insignificant for our simulation purposes.

The simulation code features a variety of preconfigured satellite constellations. Among these, a particularly notable orbit exhibits a high eccentricity, $e = 0.59$, and a perihelion distance of 0.6\,AU. This specific orbit could potentially represent a realistic experiment that might be conducted in the future. In addition to this orbit, our investigation covers a spectrum of orbital configurations, ranging from near-circular to highly elliptical trajectories. This includes orbits with semi-major axes ranging from approximately 0.1 AU to as large as about 30 AU.

The simulation software models an orbital constellation comprising four satellites. These satellites' initial state vectors are derived from a nominal state vector but are purposefully perturbed. Distances between satellites are a few thousand kilometers, with relative heliocentric velocities differing by a maximum of 0.1 m/s (see Table~\ref{tb:sim-params}).

To improve the calculation accuracy, the software adopts a moving reference frame that aligns with the satellite's nominal, unperturbed orbit. In this  frame, the satellite positions are typically within a few tens of thousands of kilometers. This limitation enables the software to achieve sub-micron level positional accuracy using standard double-precision numerical formats. This method effectively addresses the challenge of maintaining millimeter-scale accuracy in a heliocentric coordinate system, which becomes more complex when satellites are several AUs away from the Sun.

For orbit calculations, the software incorporates a fourth-order Runge-Kutta integrator. When simulating a satellite orbit at 1 AU from the Sun, it uses a 600-s timestep. This specific timestep is chosen to balance the minimization of numerical errors with the efficiency of the simulation, ensuring accurate yet expedient orbit predictions.

The main loop of the software integrates both the numerical simulation and visualization components. The visualization updates occur at a lower frequency to optimize runtime performance, ensuring smooth animation. During each iteration in the main loop:
\begin{itemize}
\item The software advances the satellite orbits using the Runge-Kutta integrator.
\item    It then integrates the reference orbit, updating the coordinate system origin for the subsequent iteration.
\item    It calculates ${\rm tr}(\vec T)$ in the inertial reference frame, serving as the ``true'' value of this trace, given the numerical constraints.
\item    Observables are derived, including the six inter-satellite ranges and the 12 generalized Sagnac observables.
\item    The software then determines the coordinates of the four vertices in the TCS using the time-series of range observables.
\item    The generalized Sagnac observables are leveraged to factor in rotation and pseudo-accelerations. The software then calculates numerical second derivatives from the coordinate time series to determine relative accelerations.
\item    These values are used to reconstruct the gravitational gradient tensor's trace in the TCS, representing the ``observed'' value. The software also factors in the second-order gravitational gradient using Eq.~(\ref{eq:grad-diff}).
 \item   These steps are replicated for all four vertices. The software then averages the "true" and "observed" gravitational gradient tensor trace values across the vertices and computes the corresponding standard deviation.
\end{itemize}

\begin{figure}[t!]
\includegraphics[]{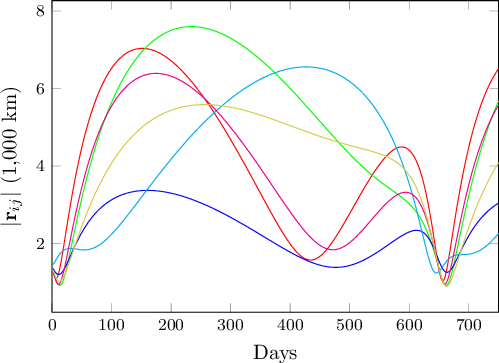}
\vskip -10pt
\caption{\label{fig:alt3} The six intersatellite ranges of the configuration introduced in Fig.~\ref{fig:alt1}.}
\vskip 6pt
\includegraphics[]{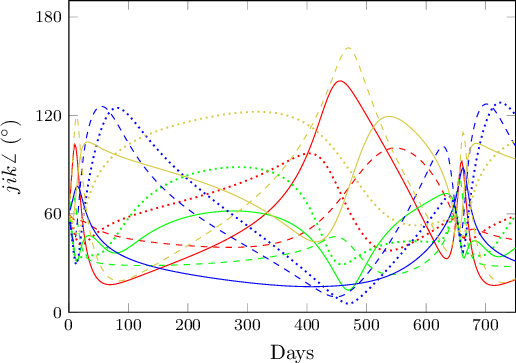}~\includegraphics[]{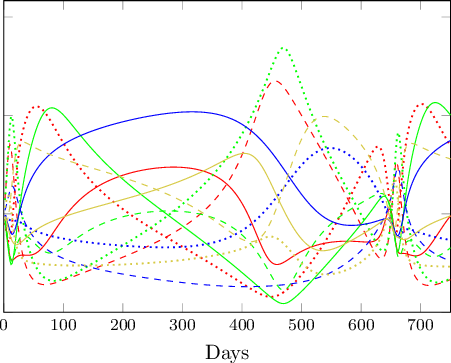}
\vskip -10pt
\caption{\label{fig:alt4} The twelve tetrahedron angles of the configuration introduced in Fig.~\ref{fig:alt1}, grouped, using color, by vertex (left) and tetrahedron face (right).
}
\end{figure}

Fig.~\ref{fig:alt1} (left) shows results for a specific orbit that has been  selected for this study with a perihelion of $\sim 0.6$~AU and an eccentricity of $e\simeq 0.5909$. In addition, the volume of the tetrahedron formed by the arrangement of the four satellites in space is depicted. This tetrahedral formation is crucial because it's the basis for the measurement methodology being discussed. When examining the data presented, it becomes evident that the majority of the computational results align well with the predicted accuracy for the trace of the GGT. However, there's a notable exception. Whenever the tetrahedron becomes ``degenerate'', meaning its three-dimensional shape collapses to a point where its volume is essentially zero, the calculations don't reproduce the expected trace of the GGT. This indicates the potential limitations or challenges of this measurement methodology under specific geometric conditions.

In contrast, Fig.~\ref{fig:alt1} (right) shows the values for the trace of the GGT that are derived from the actual ranges between satellites and the generalized Sagnac observables (see Table~\ref{tb:inst-params}). This means we're looking at data derived from more direct measurements. Additionally, to improve the accuracy of this data, the approximate distance and directional angle to the Sun, as it's perceived in the TCS (a specific observational reference), are also considered. This inclusion is necessary because there are certain gradient contributions—specifically, those of the second order—that aren't addressed by (\ref{eq:mat123d}). For clarity in interpretation, gray error bars have been added to the graph. These bars represent the variability in the values obtained when the same calculation is repeated. Each repetition is anchored at a different vertex of the tetrahedron, and there are four such vertices, leading to four repetitions.

One can see that there is a clear correlation between the geometry of the tetrahedron and the sensitivity of measurements. Specifically, when the tetrahedron's volume collapses—effectively making it flat—the ability of the satellite constellation to detect perturbations in the GGT decreases. This geometric condition hence becomes a limitation or a challenge in the study. To further elucidate this behavior, Fig.~\ref{fig:alt2} zooms into specific segments of the constellation's orbit, focusing particularly on those segments where the sensitivity to detect change in the GGT is at its maximum. This  view can help understand the conditions under which the methodology is most effective.

Additional details from this simulation are shown in Fig.~\ref{fig:alt3}, in the form of the six intersatellite ranges. We can see that the ranges change substantially during a full orbit, varying between $\sim$1,000 and $\sim$8,000~km. Such variations could be due to various factors like gravitational perturbations, inherent satellite propulsion, or design of the orbit. The range data indicate the dynamism of the satellite constellation during its operation.  Figure~\ref{fig:alt4} shows two views of the tetrahedral angles (that is to say, the angular separation of a pair of satellites as seen from a third satellite). There are twelve such angles (three per tetrahedron vertex or, alternatively, three per tetrahedron face). Correspondingly, two views are presented: in one view, the angles are grouped using the same color per vertex, whereas in the other, the grouping is by face. We can see that the tetrahedron flexes substantially, with all twelve angles changing dramatically during a full orbit. Both views reveal a significant amount of flexing in the tetrahedron throughout its orbit. The angles are not rigid but vary substantially, highlighting the tetrahedron's dynamic geometry as the satellites move.

These results are by no means unique to the specific configuration that we used for this simulation. However, this configuration performed quite well in comparison to potential alternatives. This is evident when observing the cyclical nature of the satellite constellation. After completing a full orbit, the satellites realign into a formation very close to their starting configuration. Such behavior underscores the resilience and reliability of the chosen configuration. However, the takeaways are clear: any ultimate mission that aims to use a tetrahedral configuration must account for the inherent flexibility of the tetrahedron and the variability in intersatellite distances. These factors aren't mere nuances; they are integral to the mission design. Ensuring the optimal functionality of equipment onboard, communication between satellites, and accurate data collection hinges on understanding and leveraging this dynamics.

We note that this simulation is constrained, in part, by the limits imposed by double-precision arithmetic, which yields at most $\sim 15.9$ digits of precision for simple arithmetic operations. For complex calculations, errors accumulate (random walk) so we do not expect a relative accuracy much better than $\sim 10^{-14}$. This constrains our ability to recover the ``true'' value of the trace, and further constrains our ability (in particular, by limiting the accuracy at which the Sagnac observable is modeled) to recover the ``observed'' value. Ultimately, a more accurate simulation may benefit from the use of extended precision arithmetic, which was not implemented in this prototype simulation.

In the simulation, ${\rm tr}(\vec T)$ exhibits a nonzero value with a notably tight standard deviation $\sigma_{\rm tr(T)}\lesssim 10^{-24}~{\rm s}^{-2}$. This standard deviation is computed by iterating over the four satellites, considering each as the origin point in the satellite-fixed TCS reference frame. The tightness of this standard deviation suggests that the deviation from zero is not due to random errors; rather, it indicates the presence of second-order tidal terms that have not been accounted for in the GGT. When these second-order terms (\ref{eq:grad-diff}) are included in the model, the trace of the GGT comes significantly closer to zero.  Similar improvements were observed when some of the relativistic terms were included.   

We emphasize that once the Sagnac observables are fully incorporated, the estimated ${\rm tr}(\vec T)$  value comes significantly closer to zero, offering a more accurate representation of the GGT. Therefore, the inclusion of the Sagnac-type measurement provides an important correction mechanism to improve the precision of the estimate. More broadly, in satellite constellations, determining relative accelerations requires accounting for the rotational dynamics of the reference frame. This involves determining the frame's angular velocity, achievable through the Sagnac effect in the constellation's tetrahedral configuration. By analyzing these timings bidirectionally across three satellite triplets, the frame's three-dimensional angular velocity can be accurately derived, ensuring precise acceleration measurements.

We also examined various orbits spanning from circular to eccentric and even those beyond 1 AU. We observed an enhanced accuracy in verifying ${\rm tr}(\vec T)=0$ with the expansion of orbital sizes, but significantly large heliocentric distances are nonviable. Observations highlighted intersatellite range fluctuations during orbits and accuracy reductions during tetrahedron volume collapses. The prevailing limitation was the double-precision floating-point arithmetic. For optimal orbit determination and to maximize the tetrahedral configuration's efficacy, a transition to extended precision arithmetic is imperative.

As a result, based on simulations and mission analysis, we determine that achieving a sensitivity of \(1 \times 10^{-24} \, \text{s}^{-2}\) is feasible for the satellite configuration considered. This assessment considers the precise angular velocity measurement obtained through bidirectional Sagnac effect timing across the tetrahedral satellite constellation. The targeted sensitivity reflects the technical capabilities of the system, encompassing error analysis, system response, and operational thresholds in the specified rotational dynamics context.

Finally, to evaluate the accuracy of our calculations, we introduced a modification to Newtonian gravity by incorporated a Yukawa term modeling it as: $U_{\rm mod}=(\mu_\odot/r)(1+\alpha e^{-r/\lambda})$ \cite{Adelberger:2003}. In a practical application, we conducted a sample run using Yukawa parameters $\alpha=1\times 10^{-7}$ and $\lambda=1\, {\rm AU}$. It's important to note that these parameters were chosen not to simulate any real-world modifications of gravity, but rather to test and validate the simulation code. The test run had shown that the trace corresponding to the GGT computed with $U_{\rm mod}$  and evaluated to be at a level below ${\cal O}(10^{-21}~{\rm s}^{-2})$ was detected by the constellation, showing the feasibility of the approach.  

\section{Conclusions}
\label{sec:conclude}

We have evaluated the Gravity Probe and Dark Energy Detection Mission (GDEM) -- a prospective space mission concept whose design is based on four spacecraft operating in a tight tetrahedral formation. GDEM's configuration is specifically engineered to optimize both the sensitivity and spatial resolution in precisely measuring the GGT. 

The technical feasibility of the GDEM is supported by advancements  in several key technological areas. These include spacecraft formation flying, laser interferometric ranging techniques, and the evolution of sophisticated data analysis methodologies. The implementation of GDEM relies on: 1) Precision formation flying in a tetrahedral configuration for optimal sensitivity and spatial resolution. 2) Precision laser ranging to accurately measure the distances between spacecraft. 3) Use of AI to correct for local non-gravitational disturbances and enhance the precision of gravitational measurements. 4) Sagnac interferometry is used for accurately determining the formation's angular velocity, which is essential for maintaining the tetrahedral configuration and interpreting gravitational data.

We analyzed the dynamics and behavior of a tetrahedral spacecraft formation, comprising four spacecraft in nearby elliptical heliocentric orbits. The choice of elliptical orbits over circular ones allows for sampling signals at varying heliocentric distances, thus improving detection probability. To achieve this, we developed analytical expressions that precisely describe the spacecraft dynamics within an orbital coordinate system defined  at the tetrahedron.

An important observation relates to the tetrahedron's volume evolution. Specifically, during each orbital revolution, the volume of the tetrahedron, defined by the spacecraft positions, collapses twice. This behavior is systematically represented in Fig.~\ref{fig:vol} and mathematically described by (\ref{eq:vol1}). Such changes in the volume can substantially influence the sensitivity and precision of the scientific data that are captured during the mission. We studied the evolution of the distances between the spacecraft in the formation. For orbits with an eccentricity of $e\simeq0.6$, these inter-spacecraft distances displayed notable variability. In some cases, distances expanded or contracted to levels that were up to four times their original measurements, as shown in Fig.~\ref{fig:r41}.

We also noted that prior to each of those instances when the tetrahedron's volume collapses, the quality of the solution begins to degrade. This may be addressed by resetting the tetrahedral constellation multiple times per orbit, ensuring data quality is maintained across all orbital segments. Consequently, we can gather data throughout crucial orbital segments, particularly at the apogee and perigee, which are vital in the quest for galileons.

We considered the practical implications of the tetrahedron's dynamic behavior, especially in terms of gimbal articulation. As spacecraft move in their orbits, the tetrahedron's edges change causing the relative angles between the spacecraft change by more than $80^\circ$. Such angular variations necessitate the development and deployment of gimbals that can achieve these wide articulations but also maintain stability throughout the entire angular range.

Our investigation highlighted a pivotal distinction between internal and external measurements. Traditional external references, prevalent in astrometry, exhibit limitations in precision. Given these constraints, our study advocated for the use of Sagnac observables, which are based on local measurements. This approach has the potential to surpass current methodologies, offering enhanced measurement accuracy without dependency on external reference systems.

We explored the feasibility of a tetrahedral constellation of four satellites in heliocentric orbit to precisely reconstruct the trace of the GGT. These satellites are designed to gauge inter-satellite distances and to clock the round-trip times and phases of signals coherently exchanged between them. Relying solely on inter-satellite distances poses challenges due to its ignorance on the constellation's rotational dynamics, which introduce fictitious forces. However, by integrating Sagnac-type observables, we can precisely account for and rectify these rotational discrepancies. 

Our analysis of ${\rm tr}(\vec T)$ unveiled critical second-order tidal effects overlooked in the primary formalism. If these effects predominantly arise from the Sun, then even an approximate reckoning of the Sun's position and distance in the satellite-fixed reference can considerably refine experimental accuracy. On the relativistic front, given the precision prerequisites of our experiment, a moderate understanding of the constellation's velocity relative to a heliocentric inertial frame is sufficient for calibrating on-board timekeeping mechanisms, thereby endorsing our non-relativistic modeling. To cap it off, our analysis underscores the imperative of extended precision arithmetic in models and simulations, as double precision falls short of delivering the sought-after accuracy level of ${\cal O}(10^{-24}~{\rm s}^{-2})$.

Given that the predicted force of dark energy in our solar system is vastly weaker than Newtonian gravity—by about ten orders of magnitude—GDEM's primary objective is to identify variations in the acceleration's gradient. Contrary to the zero gradient trace in Newtonian $1/r^2$ forces, other theories predict non-zero traces. GDEM targets a detection sensitivity of $10^{-24}\,{\rm s}^{-2}$ over 3-year period for these gradients. The GDEM uses the cubic galileon field to explore a potential fifth force in the solar system. Through a tetrahedral spacecraft configuration, the weak galileon force can be detected as a trace of the GGT, with measurements remaining orientation invariant.

Forces derived from $1/r$ potentials adhere to the ISL and maintain zero Laplacians in source-free regions. Unlike these, the galileon force's $1/\sqrt{r}$ dependence produces a non-zero trace corresponding to its force gradient. This allows for the direct measurement of the trace of the local GGT, circumventing gravitational inhomogeneity and eliminating the need for detailed data on mass distribution. Furthermore, the trace of GGT is a symmetric tensor is rotationally invariant, reducing concerns about the precise instrument orientation. This means the specific positioning of the measurement instrument is not a  concern, reducing potential issues related to spacecraft pointing using star trackers. 

To optimize the conditions to detect the anticipated signal that behaves  $\propto 1/\sqrt{r}$, the GDEM spacecraft will be placed on nearby elliptic heliocentric orbits that will allow to sample the galielon field at various distances from the Sun. Based on our simulations of the galielon field, there is an order of magnitude of the galielon force variation in the solar system. An elliptical orbit with varying distance from the Sun will allow the observation of such variation. This distance-dependent variation would significantly reduce the systematics yielding a stronger evidence for a GR violation, if observed. This insight will be used to further assess the relevant mission and instrument requirements. 

As a result, we have shown that the tetrahedral satellite constellations offer a promising avenue for precision gravitational measurements. In the quest to accurately measure variations in the gravitational field, particularly the trace ${\rm tr}(\vec T)$, the configuration's potential becomes evident. Using the data about the Sun's position and distance, our system—comprising satellites typically spaced 1,000 km apart and orbiting with a semi-major axis of 1 AU—shows capability to achieve a measurement precision approaching ${\cal O}(10^{-24}~{\rm s}^{-2})$. Such precision provides a tangible means to probe for galileonic deviations in the solar gravitational field, potentially significantly improving our current understanding. More broadly, the mission obtained data set may also be used for other science analysis including dark matter detection and detection of gravitational waves  within the so-called mid-band frequency \cite{Bailes-etal:2021}.

To conclude, the Gravity Probe and Dark Energy Detection Mission (GDEM) mission is undeniably ambitious, yet our analysis underscores its feasibility within the scope of present and emerging technologies. In fact, the key technologies required for GDEM, including precision laser ranging systems, atom-wave interferometers, and Sagnac interferometers, either already exist or are in active development, promising a high degree of technical readiness and reliability. A significant scientific driver for the GDEM lies in the potential to unveil non-Einsteinian gravitational physics within our solar system—a discovery that would compel a reassessment of prevailing gravitational paradigms.  If realized, this mission would not only shed light on the nature of dark energy but also provide critical data for testing modern relativistic gravity theories. It has the potential to advance the search for ultra-light fields of dark matter and facilitate gravitational wave detections in the mid-band frequency spectrum. This paper sets forth the requisite technological and methodological foundations essential to the GDEM's successful execution. While this constitutes a significant stride, the relevant work continues, and subsequent findings will be communicated in future publications.

\begin{acknowledgments}

We would like to express our gratitude to our many colleagues who have either collaborated with us on this manuscript or given us their wisdom. We specifically thank all the NIAC team members at JPL, including Jason D. Rhodes, Olivier P. Dore,  Jeffrey B. Jewell,  Curt J. Cutler, Rashied Amini, and Kristofer M. Pardo, who benefited us with valuable discussions, insightful comments and important suggestions. SGT acknowledges very fruitful collaboration with Viktor T.Toth who provided many valuable comments and participated in stimulating discussions on the various topics discussed in this document. Contributions and discussions in formulating the GDEM mission from the entire study team are acknowledged. The study was sponsored under the NASA NIAC program. The work described here was carried out at the Jet Propulsion Laboratory, California Institute of Technology, under a contract with the National Aeronautics and Space Administration.
\textcopyright 2024. California Institute of Technology. Government sponsorship acknowledged.

\end{acknowledgments}


\appendix

\section{Relevant partial derivatives}
\label{sec:partials}

The partials of $r$ and $\nu$ with respect to $\alpha$ are computed by first taking the partials of $M$ with respect to $\alpha$,
{}
\begin{eqnarray}
M = M_0 + n(t-t_0)\quad
&\Rightarrow&
\quad
\frac{\partial M}{\partial e}
= \frac{\partial M}{\partial i}
= \frac{\partial M}{\partial \Omega}= \frac{\partial M}{\partial \omega} = 0,
\qquad
\frac{\partial M}{\partial a}
=-\frac{3n}{2a}(t -t_0), \qquad \frac{\partial M}{\partial M_0}= 1.
\end{eqnarray}
The partials of $M$ are then related to the partials of the eccentric anomaly, $E$, through Kepler's equation, from $M = E - e \sin E$, we derive
{}
\begin{align}
\frac{\partial E}{\partial i}
= \frac{\partial E}{\partial \Omega}
= \frac{\partial E}{\partial \omega}= 0,
\quad
\frac{\partial E}{\partial a}
=-\frac{3n}{2r}(t - t_0),
\quad \frac{\partial E}{\partial e}= \frac{\sin E}{1 - e
\cos E} = \frac{\sin \nu}{\sqrt{1 - e^2}}, 
\qquad \frac{\partial E}{\partial M_0}
=\frac{1}{1 - e \cos E}=\frac{a}{r}.
\end{align}

The partials of $r$ can be related to the partials of $E$ through several
equations. From $r = a\big(1- e \cos E\big)$, we have:
{}
\begin{align}
\frac{\partial r}{\partial i}
= \frac{\partial r}{\partial \Omega}
= \frac{\partial r}{\partial \omega}= 0, 
\qquad
\frac{\partial r}{\partial a}
=\frac{r}{a}-\frac{3n(t - t_0)e\sin\nu}{2\sqrt{1-e^2}},
\qquad
\frac{\partial r}{\partial e}= -a \cos \nu, 
\qquad 
\frac{\partial r}{\partial M_0}
=\frac{ae\sin\nu}{\sqrt{1-e^2}}.
\end{align}
Also, the partials of $\nu$ can be related to the partials of $r$ and $E$ through the equation $a\cos E= ae +r \cos \nu$, yielding
{}
\begin{align}
\quad \frac{\partial \nu}{\partial i}
= \frac{\partial \nu}{\partial \Omega}
= \frac{\partial \nu}{\partial \omega}= 0, 
\quad \frac{\partial \nu}{\partial a}
=-\frac{3a}{2r^2}n(t - t_0)e\sqrt{1-e^2},\quad
\frac{\partial \nu}{\partial e}= \frac{\sin\nu}{1-e^2}\Big(2+e\cos\nu\Big), \quad \frac{\partial \nu}{\partial M_0}
=\frac{a^2}{r}\sqrt{1-e^2}.
\end{align}

The partial derivatives of $\vec r$ with respect to $\alpha$ are computed using the preceding results for $\partial \vec r/ \partial \alpha$ and $\partial \nu/ \partial \alpha$. For that, introducing
{}
\begin{eqnarray}
\vec \Pi_1 &=&
   \Bigg[
 \begin{aligned}
\cos\Omega\cos\theta&-\sin\Omega\cos i \sin\theta \\
\sin\Omega \cos\theta&-\cos\Omega\cos i\sin\theta\\
&\sin i \sin\theta
  \end{aligned}\Bigg],
  \qquad
  \vec \Pi_2 =
   \Bigg[
 \begin{aligned}[c]
-\cos\Omega\sin\theta&-\sin\Omega\cos i \cos\theta \\
-\sin\Omega \sin\theta&+\cos\Omega\cos i\cos\theta\\
&\sin i \cos\theta
  \end{aligned}\Bigg],
  \label{eq:r-vec2}
\end{eqnarray}
we have:
{}
\begin{eqnarray}
\frac{\partial \vec r}{\partial a}&=&\frac{\partial  r}{\partial a}
\vec \Pi_1
  +r\frac{\partial \nu}{\partial a}
\vec \Pi_2,
  \label{eq:r-vec6}
  \qquad
 \frac{\partial \vec r}{\partial e}=\frac{\partial  r}{\partial e}\vec \Pi_1 +r\frac{\partial \nu}{\partial e}\vec \Pi_2, \qquad
 \frac{\partial \vec r}{\partial M_0}=\frac{\partial  r}{\partial M_0}\vec \Pi_1 +r\frac{\partial \nu}{\partial M_0}\vec \Pi_2,
\end{eqnarray}
{}
\begin{eqnarray}
\frac{\partial \vec r}{\partial i}&=&r\Bigg[
 \begin{aligned}
\sin\Omega\sin i \sin\theta \\
-\cos\Omega\sin i\sin\theta\\
\cos i \sin\theta
  \end{aligned}\Bigg],
\qquad
 \frac{\partial \vec r}{\partial \Omega} =
 r  \Bigg[
 \begin{aligned}
-\sin\Omega\cos\theta&-\cos\Omega\cos i \sin\theta \\
\cos\Omega \cos\theta&-\sin\Omega\cos i\sin\theta\\
&0
  \end{aligned}\Bigg],
  \qquad
 \frac{\partial \vec r}{\partial \omega}  = r
\vec \Pi_2.
  \label{eq:r-vec4}
\end{eqnarray}

Thus, the partials of $\vec r$ with respect to $\alpha$ can be written as the sensitivity
matrix $S_{\rm HCI}$,
\begin{eqnarray}
S_{\rm HCI}&=&\bigg[
 \begin{aligned}
\frac{\partial \vec r}{\partial a}\Big|
\frac{\partial \vec r}{\partial e}\Big|
\frac{\partial \vec r}{\partial i}\Big|
\frac{\partial \vec r}{\partial \Omega}\Big|
\frac{\partial \vec r}{\partial \omega}\Big|
\frac{\partial \vec r}{\partial M_0}
  \end{aligned}\bigg]_{3\times 6}.
  \label{eq:r-vec5}
\end{eqnarray}

\end{document}